\documentclass[10pt,twocolumn,floatfix,pra]{revtex4-1}

\usepackage{graphicx}
\usepackage{epsfig}
\usepackage{dcolumn}
\usepackage{bm}
\usepackage{amsmath}
\usepackage{amssymb}
\usepackage{graphicx}
\usepackage{esint}
\usepackage{color}

\newcommand{\uu}[1]{{\boldsymbol #1}}

\def\pb{\uu{p}}
\def\xb{\uu{x}}
\def\nb{\uu{n}}
\def\etab{\uu{\eta}}
\def\<{\langle}
\def\>{\rangle}

\def\Deltai{\mathit{\Delta}}

\DeclareMathOperator{\argmax}{argmax}

\def\RR{\mathbb{R}}

\begin{document}

\title{Optimized Markov State Models for Metastable Systems}

\author{Enrico Guarnera}

\altaffiliation{Current Address: Bioinformatics Institute (BII), Agency for Science, Technology and Research (ASTAR), Singapore}
\email{enricog@bii.a-star.edu.sg}
\affiliation{%
  Courant Institute of Mathematical Sciences, \\
New York University, \\
New  York, NY 10012, USA
}%
\author{Eric Vanden-Eijnden}
\email{eve2@cims.nyu.edu}
\affiliation{%
  Courant Institute of Mathematical Sciences, \\
New York University, \\
New  York, NY 10012, USA
}%

\begin{abstract}
  A method is proposed to identify target states that optimize a
  metastability index amongst a set of trial states and use these
  target states as milestones (or core sets) to build Markov
  State Models (MSMs). If the optimized metastability index is
  small, this automatically guarantees the accuracy of the MSM, in the
  sense that the transitions between the target milestones is indeed
  approximately Markovian. The method is simple to implement and use,
  it does not require that the dynamics on the trial milestones be
  Markovian, and it also offers the possibility to partition the
  system's state-space by assigning every trial milestone to the
  target milestones it is most likely to visit next and to identify
  transition state regions. Here the method is tested on the
  Gly-Ala-Gly peptide, where it shown to correctly identify the expected
  metastable states in the dihedral angle space of the molecule
  without \textit{a~priori} information about these states. It is also
  applied to analyze the folding landscape of the Beta3s mini-protein,
  where it is shown to identify the folded basin as a connecting hub
  between an helix-rich region, which is entropically stabilized,
    and a beta-rich region, which is energetically stabilized and acts
    as a kinetic trap.
\end{abstract}


\maketitle

\section{Introduction}
\label{sec:intro}

Markov State Models (MSMs) have become an integral part of the toolbox
used to analyze the output of molecular dynamics (MD) simulations of
complex systems such as proteins and other large
biomolecules~\cite{Schutte:1999ua,Swope:2004ab,Chodera:2007aa,%
  Noe:2008aa,Schutte:2011aa,Schutte:2013aa,Bowman:2013aa}.  They were
developed in response to the need to process ever longer MD timeseries
data, made either of one long trajectory or very many shorter ones,
generated e.g by special-purpose high-performance
computers~\cite{Shaw:2010aa}, high-performance
GPUs~\cite{Stone:2007aa}, or massively parallel
simulations~\cite{Voelz:2010aa}.  The basic idea of MSMs is to
represent the original dynamics as memoryless jumps between predefined
states in the configuration space of the molecular system. Under this
Markovian assumption, MD timeseries data can then be processed via
inference techniques such as maximum likelihood estimation to
calculate the rate matrix between these states. This matrix defines a
Markov jump process (MJP), which in turns permits the calculation of
interesting kinetic quantities of the system on time scales that may
be larger than those reached in the MD simulations: indeed MSMs also
permit to recombine short simulations run in parallel to extract long
time information about the system.

A recurrent issue in the context of MSMs is how to pick the states on
which to map the original dynamics -- see
  Refs.~\cite{McGibbon:2014,Sultan:2014,Martini:2016} for some recent
  works in this direction. Indeed this mapping amounts to a drastic
coarse-graining of the dynamics, and the jumps between poorly
  chosen coarse states will not be Markovian in general. This
  invalidates the basic assumption of MSMs and affect their
  reliability and accuracy. Fortunately, MD systems typically display
metastability and this offers a way around this difficulty. In
metastable systems there exists regions that play the role of hubs:
after visiting one such hub, the system returns often to it before
making a transition to another. This guarantees that transitions
between these hubs is indeed approximately Markovian, and
  metastability has therefore been invoked as they key property to
  justify MSMs and assess their accuracy (see
  e.g.~[\onlinecite{Schutte:2013aa}] and [\onlinecite{,Bowman:2013aa}]
  for modern perspectives on the topic that summarizes this
  viewpoint). What remains mostly open, however, is how to identify
  these hubs in practice.

In the present paper we aim at addressing this question in the context
of milestoning-based MSMs \cite{Schutte:2011aa} that combine the
core set method originally introduced in
Ref.~[\onlinecite{Buchete:2008vn}] with the milestoning of the
trajectories developed in Refs.~[\onlinecite{Faradjian:2004aa,%
  Vanden-Eijnden:2008aa,Vanden-Eijnden:2009ab}]. Unlike standard MSMs
that are based on a full partition of the configuration space of the
system into blocks that are used as
states~\cite{Krivov:2006aa,Chodera:2007aa,%
  Sarich:2010aa,Beauchamp:2011aa,Vitalis:2012uq,Fackovec:2015},
milestoning-based MSMs uses non-adjacent core sets as states (the
  milestones), and assign the MD timeseries to the index of the last
such milestone it visited. This maps the original dynamics onto a
symbolic dynamics on these indices that is then used as input to build
the MSM by maximum likelihood estimation of its transition matrix. In
metastable systems, the proper milestones to use should be the hubs
mentioned before. Here we propose to identify these hubs among a set
of trial milestones via optimization of a metastability index that
measures how good the hubs are. The method can be justified within the
framework of the potential theoretic approach to metastability
developed by Bovier and
collaborators~\cite{Bovier:2002ab,Bovier:2004aa,Bovier:2005aa}. It has
the advantage that it can be used even in situations where the
dynamics on the trial milestones is non-Markovian. In this sense, it
alleviates a difficulty with the standard approach used to build MSMs
via clustering of trial states~\cite{Bowman:2013aa}: This clustering
is typically done using spectral analysis of the rate matrix of the
chain built on these trial states, which may lead to artifacts since
the dynamics on these trial states is non-Markovian in general. The
method we propose avoids this difficulty altogether. In addition
  it avoids the need to introduce a time-lag to process the data,
  which may be difficult to adjust. As we will see, our method also
offers a way to partition the state space of the system by identifying
regions made of configurations most likely to reach a given hub,
  and to identify the members of the transition state ensemble as
  those trial milestones that have a non-negligible probability to
  reach more than one target milestone next.

The remainder of this paper is organized as follows. In
Sec.~\ref{sec:algo} we start by presenting the algorithmic aspects of
the method we propose, including how to define the trial milestones
(Sec.~\ref{sec:trialvstarget}) and the metastability index
(Sec.~\ref{sec:metaindex}), how to identify the target milestones that
optimize this metastability index (Sec.~\ref{sub:How-to-find}), and
how to build the MSM on these target milestones
(Sec.~\ref{sec:MSM}). We also introduce a variety of diagnostic tests
that can be used \textit{a~posteriori} to analyze the output of the
MSM and use the trial milestones to get additional information about
the system's dynamics (Sec.~\ref{sec:diagno}). A theoretical
justification of this algorithm is then given in Sec.~\ref{sec:theo},
first in the context of Markov jump processes, which is relevant
e.g. if one assumes that the dynamics on the trial milestones is
itself Markovian (Sec.~\ref{sec:ctmc}), then in the context of systems
whose configurational space is continuous, like those encountered in
MD simulations, where we cannot expect the dynamics on the trial
milestones to be Markovian (Sec.~\ref{sec:csp}).  We also test the
method on a simple one-dimensional example with a multiscale energy
landscape (Sec.~\ref{sec:illustrativeex}). In Sec.~\ref{sec:glyalagly}
we then apply the method to analyze the dynamics of a Gly-Ala-Gly
peptide, and in Sec.~\ref{sec:Beta3s} we use it to analyze the folding
pathways of the Beta3s mini-protein. Concluding remarks are given in
Sec.~\ref{sec:conclu}. Some technical derivations are relegated to an
Appendix.

\section{Algorithmic aspects}
\label{sec:algo}

In this section, we outline the algorithm we propose to identify the
milestones (or core sets) over which to build an MSM. In a
nutshell, this is done by picking among a set of trial milestones a
subset of target milestones which minimizes a metastability index --
this index is defined so that small values are indicative of
Markovianity. The target milestones are taken as states in the MSM,
while the trial milestones are used to complement the predictions of
this MSM and e.g. partition the system's configurational space or
identify transition state regions in it.

\begin{figure}[t]
\begin{centering}
\includegraphics[scale=0.35]{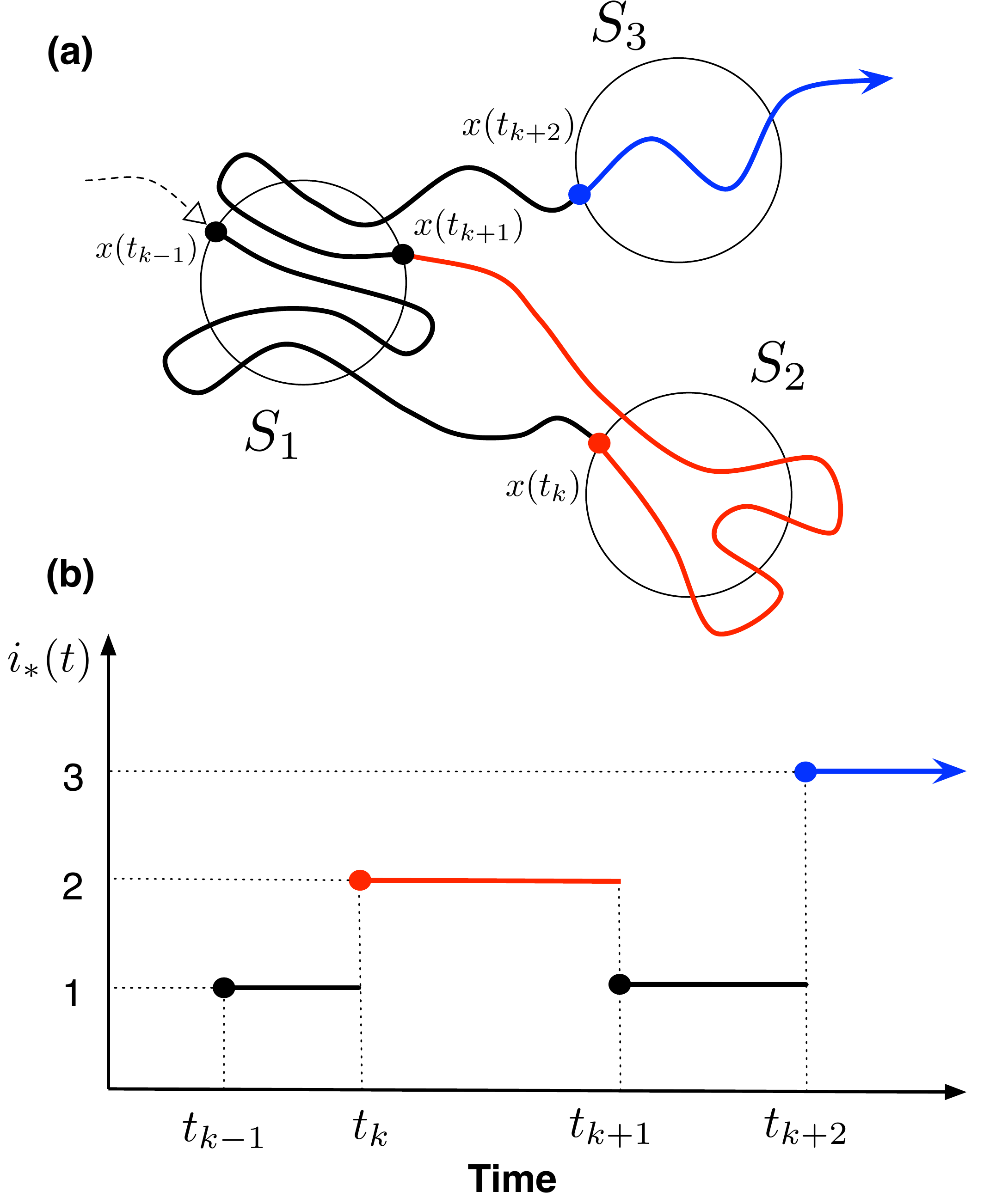}
\end{centering}
\caption{\label{fig:milestoning} Milestoning procedure. (a):
  piece of a long trajectory $\xb(t)$ crossing a set of three circular
  milestones, $S_{1}$, $S_{2}$, and $S_{3}$. (b): the trajectory
  shown in (a) is mapped onto the index of the last milestone it
  hit, thereby defining the piecewise constant function $i_{*}(t)$.}
\end{figure}

\subsection{Trial versus target milestones}
\label{sec:trialvstarget}

Denote by $\xb(t) \in \RR^{3n}$ a trajectory containing the
instantaneous position of the $n$ atoms in a molecular system. We
assume that we have generated one or several such trajectories
  and our goal is to build a MSM that captures their main features.
To this end, we introduce a set of~$N$ trial milestones, which we will
denote by $S_1$, $S_2$, \ldots, $S_N$ and label them by their index,
i.e.~$i$ identifies~$S_i$. These milestones are disjoint sets in the
system's configuration space that can be defined e.g. by requiring
that some of the dihedral angles of the molecules take values between
specific bounds, etc. -- how to actually choose the trial
milestones~$S_i$ will be illustrated below on specific examples.  In
the spirit of milestoning, we then map each trajectory $\xb(t)$ onto
the index of the last trial milestone $S_i$ it hit, see
Fig.~\ref{fig:milestoning} for an illustration. This way we obtain a
piecewise constant index function $i_*(t)$ whose value at time $t$ is
the index of the last milestone hit by~$\xb(t)$.  Note in this
procedure we discount recrossings: we only update the index function
when a new milestone is hit.

We stress that at this stage we do not assume that the dynamics of the
index function~$i_*(t)$ on the trial milestones is Markov -- in general
it will not be. What we would like to do next is extract from the set
of $N$ trial milestones a subset of $M\le N$ target milestones such
that if we map the trajectory~$\xb(t)$ onto this subset of target
milestones, the corresponding index function will be approximately
Markovian -- these target milestones are shown in red in the cartoon
shown in Fig.~\ref{fig:metastabilitySCHEME}. In the sequel, we will
denote by $\mathcal{M} = \{i_1, i_2, \ldots, i_M\}\subset
\{1,2,\ldots, N\}$ the set of indices identifying the target
milestones, i.e. these are $\{S_{i_1}, S_{i_2}, \ldots, S_{i_M}\}$. We
will also refer to trial milestones that are not target ones as
non-target milestones.

\subsection{Metastability index}
\label{sec:metaindex}

\begin{figure}
\begin{centering}
\includegraphics[scale=0.5]{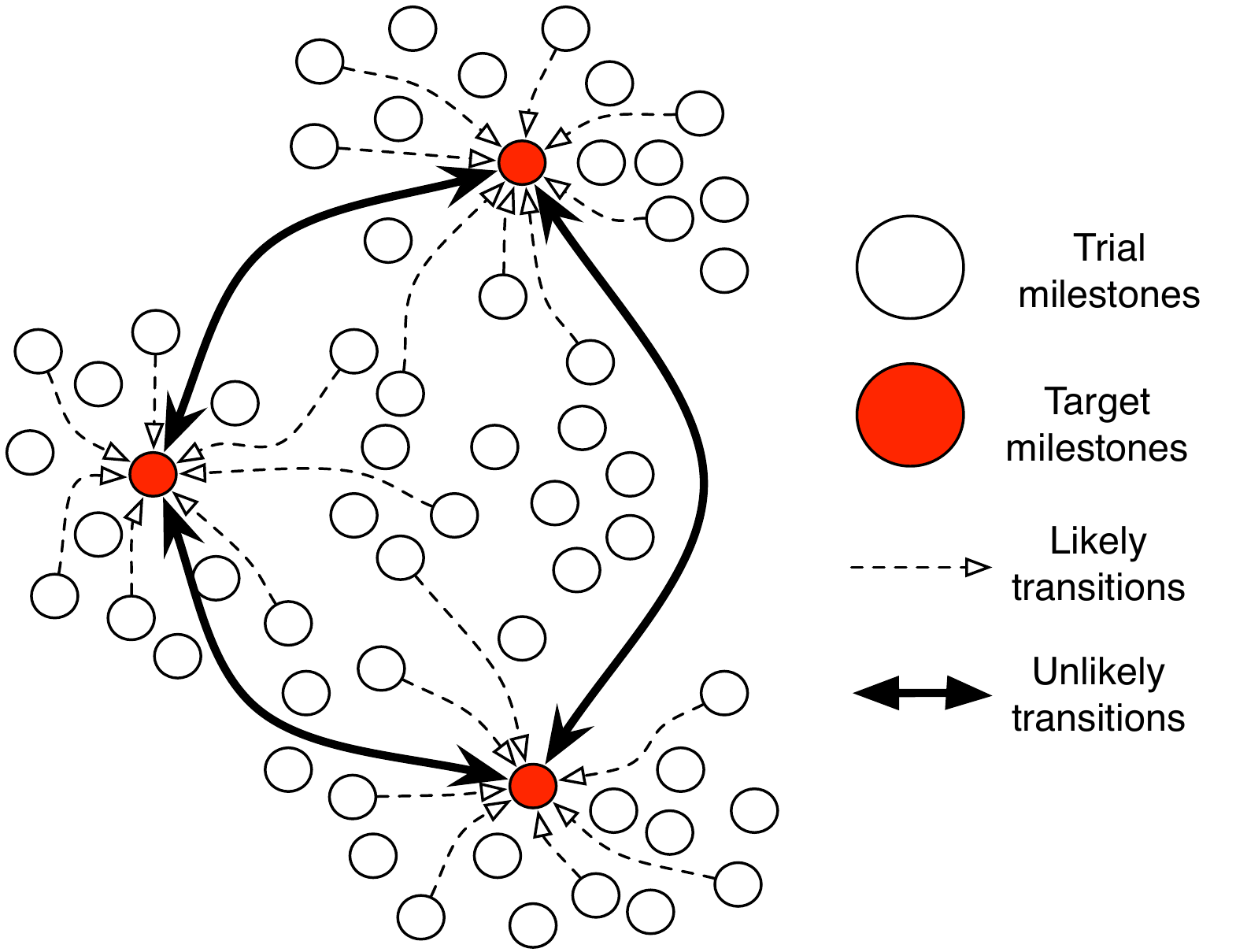}
\end{centering}
\caption{\label{fig:metastabilitySCHEME}Schematic representation of a
  subset of target milestones (shown in red) immersed in a set of
  trial milestones. A good subset of target milestone
  is such that transitions from non-target to target milestones
  are likely, while transitions between target milestones are
  not. This can be quantified via the metastability index defined
  in~\eqref{eq:6}.}
\end{figure}

How should the target milestones be chosen? Intuitively, they should
be such that they are hubs among the trial milestones towards which
the trajectory is attracted but between which it seldom makes
transitions, as illustrated in Fig.~\ref{fig:metastabilitySCHEME}. To
make this idea concrete, let us first introduce the probability
$\Gamma_{\! i,j}$ that, if the trajectory hits milestone $S_i$, then
subsequently it will hit $S_j$ with $j\not = i$ before hitting $S_i$
again (discounting recrossings: recall that only hits of different
milestones are counted -- in other words, to hit $S_i$ again, the
trajectory must have hit at least one other milestone in between). To
estimate $\Gamma_{\! i,j}$, out of each of the piecewise constant index function
$i_*(t)$, we first extract the sequence $\{i_1,i_2,i_3,..\}$ of
successive values that this function takes -- for example, for the
timeseries illustrated in Fig.~\ref{fig:milestoning} this sequence
starts with $i_1=1$, $i_2=2$, $i_3=1$, $i_4=3$, etc. We then cut this
sequence into the $N_i$ pieces which start from $i$ and contain all
the indices visited after $i$ before $i$ appears again. For example,
if the sequence is made of 3 indices, $i,j,k$, and reads
\begin{equation}
  \label{eq:3}
  \{i,k,i,j,k,j,i,k,j\},
\end{equation}
we cut it into three pieces ($N_i=3$)
\begin{equation}
  \label{eq:4}
  \{i,k\}; \quad \{i,j,k,j\}; \quad \text{and} \quad \{i,k,j\}
\end{equation}
Finally we count the number of pieces $N_{i,j}$ in which $j$ appears
at least once (in the example above $N_{i,j}=2$ since $j$ appears in
the last two pieces but not in the first), we add up these
  numbers coming from every piece of $i_*(t)$ at our disposal, and we
  set
\begin{equation}
  \label{eq:2}
  \hat\Gamma_{\! i,j} =\frac{N_{i,j}}{N_i}
\end{equation}
as estimator for $\Gamma_{i,j}$. Note that the quality of this
estimator depends on the lengths of the pieces of timeseries $\xb(t)$
that we have at our disposal, and how to assess the statistical
accuracy of~\eqref{eq:2} is nontrivial. As usual, we cannot expect the
estimator to be accurate if these pieces are too short to observe all
the relevant events in the dynamics of the system: In the context
of~\eqref{eq:2} this requires that these pieces be long enough that
trajectories starting at a non-target milestone have time to reach a
target one.  Indeed, as we will see below, this condition is
sufficient to guarantee that we will correctly identify target from
non-target milestones among the trial ones.

We will now use the matrix with entries $\Gamma_{\! i,j}$ to quantify how
good a set of target milestones will be to build an MSM. Specifically,
given a candidate $\mathcal{M} = \{i_1, i_2, \ldots, i_M\}$
identifying the target milestones, we estimate the quality of these
milestones via their metastability index defined as
\begin{equation}
  \label{eq:6}
  \rho_{\mathcal{M}} = \frac{\max_{i\in \mathcal{M}}\max_{j\in \mathcal{M}
      \setminus \{i\}} \Gamma_{\! i,j}}
  {\min_{i\not \in \mathcal{M}}\max_{j\in \mathcal{M}} \Gamma_{\! i,j}}
\end{equation}
The smaller $\rho_{\mathcal{M}}$, the better the set of target
milestones identified by $\mathcal{M} = \{i_1, i_2, \ldots,
i_M\}$. This claim will be justified in Sec.~\ref{sec:theo} by
connecting $ \rho_{\mathcal{M}}$ to a quantity originally introduced
by Bovier, but let us briefly explain here why it is true. The
numerator in~\eqref{eq:6},
\begin{equation}
  \label{eq:7}
  \max_{i\in \mathcal{M}}\max_{j\in \mathcal{M}
      \setminus \{i\}} \Gamma_{\! i,j},
\end{equation}
identifies the target milestone $S_i$ for which the probability is the
highest that, after hitting $S_i$, the trajectory will hit some other
target milestone $S_j$ before hitting $S_i$ again. In this sense,
$S_i$ is the worst target milestone in the set since we would like
that transitions between these target milestones be unlikely, and the
smaller \eqref{eq:7} the better. Correspondingly, the denominator
in~\eqref{eq:6},
\begin{equation}
  \label{eq:8}
  \min_{i\not \in \mathcal{M}}\max_{j\in \mathcal{M}} \Gamma_{\! i,j},
\end{equation}
identifies the non-target milestone $S_i$ which is such that the
trajectory has the lowest probability to hit a target milestone $S_j$
before hitting $S_i$ again. In this sense, $S_i$ is the worst
non-target milestone since we would like that transitions from
non-target to target milestones be likely, and the larger \eqref{eq:8}
the better. The metastability index $\rho_{\mathcal {M}}$
in~\eqref{eq:6} accounts for both the desiderata that \eqref{eq:7} be
small and~\eqref{eq:8} be large, and in this sense it measures the
quality of the set of target milestones identified by $\mathcal{M} =
\{i_1, i_2, \ldots, i_M\}$.

\subsection{Target milestones identification}
\label{sub:How-to-find}

Since good sets of target milestones are those whose metastability
index $\rho_{\mathcal{M}}$ is small, we can systematically search for
such good sets by minimizing $\rho_{\mathcal{M}}$.  In principle, this
can be done by considering increasing values $M=2$, $M=3$, etc. of the
cardinal of $\mathcal{M}$, and for each compute $\rho_{\mathcal{M}}$
for every choice of $\mathcal{M} = \{i_1, i_2, \ldots, i_M\}$ so as to
identify the one with minimum $\rho_{\mathcal{M}}$. Any choice
$\mathcal{M}$ for which $\rho_{\mathcal{M}}$ is small leads to a good
set of target milestones. Note however that $\rho_{\mathcal{M}}$ can
be small for different values of $M$, and so several of them should be
considered -- this effect will be illustrated in the examples
below. Of course, if the number $N$ of trial milestones is large, this
direct search strategy will quickly become inefficient as $M$
increases since the number of sets $\mathcal{M}$ to consider for each
$M$ is
\begin{equation}
  \label{eq:1}
  \frac{N!}{M!(N-M)!}
\end{equation}
To avoid this difficulty, we must adopt more efficient optimization
strategies, for example using  Monte Carlo schemes.  We have used such
schemes in the examples below. However, we found that the following
searching strategy was typically the most efficient. 

Given $\Gamma_{\! i,j}$, we can identify the index $j^\dagger (i)$ of the milestone that
the trajectory is most likely to hit after hitting~$i$ as the one that
maximizes $\Gamma_{\! i,j}$ over all $j\not = i$, i.e.
\begin{equation}
  \label{eq:9}
  \Gamma_{i, j^\dagger (i)} = \max_{k\not=i} \Gamma_{i, k}
\end{equation}
The function $j^\dagger (i)$ can be used to define `trajectories' in
index space: given $i$, we update it to $j^\dagger (i)$, then to
$j^\dagger (j^\dagger (i))$, etc. Correspondingly, given a set
$\mathcal{M}=\{i_1,i_2, \ldots, i_M\}$ we can update this set by
updating every entry in it, and only keeping the ones that remain
different (for example we could have $j^\dagger (i_1)= j^\dagger
(i_2)$ in which case only one of these entries is kept in the update
of $\mathcal{M}$). Because this updating identifies set of indices of
milestones that are likely to be hit, these sets should have small
metastability index $\rho_{\mathcal{M}}$, and this is indeed what we
observed in practice. Specifically, we took random sets of indices
$\mathcal{M}=\{i_1,i_2, \ldots, i_M\}$ with random values of $M$ and
updated them as described above while monitoring the metastability
index $\rho_{\mathcal{M}}$ of these updated sets. We observed that
this metastability index typically diminishes before starting to
oscillates in a periodic fashion (this is because the update has no
fixed point, $j^\dagger (i) \not = i$ by construction). When this
happened we stopped the update, kept the updated set
$\mathcal{M}=\{i_1,i_2, \ldots, i_M\}$ with smallest
$\rho_{\mathcal{M}}$, and restarted the procedure with a different
random set of indices $\mathcal{M}=\{i_1,i_2, \ldots, i_M\}$.  After a
few such iterations, we typically got a few different sets
$\mathcal{M}=\{i_1,i_2, \ldots, i_M\}$ (with different $M$) with small
$\rho_{\mathcal{M}}$.

\subsection{MSM building on target milestones}
\label{sec:MSM}

Once we have identified a good set of target milestones specified by
the index set $\mathcal{M}=\{i_1,i_2, \ldots, i_M\}$, we can build an
MSM using these milestones as states. How to do so was explained
  in Refs.~\cite{Vanden-Eijnden:2009ab,Schutte:2011aa}, so let us be
  brief here and refer the reader to the original papers for details.
For the sake of clarity, in the sequel it will be convenient to
distinguish between target and non-target milestones: we will do so by
using greek letters $\alpha$, $\beta$, ... to refer to the indices in
the index set $\mathcal{M}$, and $B_\alpha$, $B_\beta$, ... to refer
to the target milestones $S_{i_1}$, $S_{i_2}$, ...

Similarly to what was done before, we can map the trajectory $\xb(t)$
onto the index of the last target milestone it hit. This defines a
piecewise constant function $\alpha_*(t)$ taking values in
$\mathcal{M}$. Because we are now using target milestones instead of
trial ones, unlike $i_*(t)$, $\alpha_*(t)$ should be approximately
Markov. This means that the evolution of this function can be
completely specified by a rate matrix with entries $k_{\alpha,\beta}$:
to leading order in $\delta t \ll1$, $k_{\alpha,\beta} \delta t$ gives
the probability that $\alpha_*(t)$ jumps from the value $\alpha$ to
the value $\beta\not=\alpha$ in the interval $[t,t+\delta t)$
(i.e. that $\xb(t)$ hits the target milestone $B_\beta$ in that time
interval if the last target milestone it hit before time $t$ was
$B_\alpha$). In particular, if $p_\alpha(t)$ denotes the probability
distribution that $\alpha_*(t)$ takes the value $\alpha$ at time~$t$,
then $p_\alpha(t)$ satisfies the master equation
\begin{equation}
  \label{eq:10}
  \frac{dp_\alpha(t)}{dt} = \sum_{\beta \not= \alpha} \left(p_\beta(t)
    k_{\beta,\alpha} - p_\alpha(t) k_{\alpha,\beta}\right)
\end{equation}
Similarly, we can write down equations for the distribution of first
passage time from target milestone $B_\alpha$ to target milestone
$B_\beta$, its mean, etc. in terms of $k_{\alpha,\beta}$ -- see
e.g. Refs.~\cite{Vanden-Eijnden:2009ab,Schutte:2011aa}.

The rate matrix entries $k_{\alpha,\beta}$ can be estimated from the
timeseries $\xb(t)$ by the method of maximum likelihood. If we denote
by $T_\alpha$ the total time the last target milestone hit by $\xb(t)$
is $B_\alpha$ and by $N_{\alpha,\beta}$ the number of times the target
milestone $B_\beta$ was hit directly after $B_\alpha$ along this
timeseries, the maximum likelihood estimator for $k_{\alpha,\beta}$ is
\begin{equation}
  \label{eq:5}
  \hat k_{\alpha,\beta} = \frac{N_{\alpha,\beta}}{T_\alpha}
\end{equation}
This estimator is unbiased in the sense that, if $\alpha_*(t)$ is
indeed Markov with rate matrix entries $k_{\alpha,\beta}$ and the
length of the timeseries tends to infinity, then
$\hat k_{\alpha,\beta} \to k_{\alpha,\beta}$ in this limit. If the
length of the timeseries is finite, the statistical errors on
$\hat k_{\alpha,\beta}$ can be estimated by Bayesian sampling, see
Ref.~[\onlinecite{Schutte:2011aa}] for detail. Another source of
  errors are those due to residual non-Markovian effects in
  $\alpha_*(t)$. The smaller $\rho_{\mathcal{M}}$, the smaller these
  non-Markovian effects are, as discussed in Sec.~\ref{sec:theo}. In
  practice they can also be estimated via Markovianity tests, as
  illustrated on examples in Secs.~\ref{sec:glyalagly}
  and~\ref{sec:Beta3s}.  Notice also that if $\xb(t)$ satisfies
detailed balance, then we should have that
$N_{\alpha,\beta}/N_{\beta,\alpha}\to1$ as the length of the
timeseries goes to infinity. When this length is finite, however,
$N_{\alpha,\beta} \not= N_{\beta,\alpha}$ in general, and to enforce
detailed balance of the MSM, it is better to use the following
symmetrized estimator for the rate matrix entries:
\begin{equation}
  \label{eq:5b}  \hat k_{\alpha,\beta} = \frac{N_{\alpha,\beta}+N_{\beta,\alpha}}{2T_\alpha}
\end{equation}
With this choice, the equilibrium distribution of the MSM, i.e. the
distribution $\hat \pi_\alpha$ towards which the solution
to~\eqref{eq:10} with $k_{\alpha,\beta}$ replaced by its estimator
$\hat k_{\alpha,\beta}$ converges as $t\to\infty$, is
\begin{equation}
  \label{eq:13}
  \hat\pi_\alpha = \frac{T_\alpha}{T}
\end{equation}
where $T=\sum_\alpha T_\alpha$ is the total length of the timeseries
used to estimate $\hat k_{\alpha,\beta}$.  The distribution
  $\hat\pi_\alpha$ gives the proportion of time during which $B_\alpha$
  was the last milestone hit by $\xb(t)$ and as the length of the
  timeseries increases, $T\to\infty$, $\hat \pi_\alpha$ converges to the
  true equilibrium distribution of milestone $B_\alpha$ -- explicit
  expressions for this distribution in the context of a system whose
  dynamics is governed by a Markov jump process or a Langevin equation
  will be given in Secs.~\ref{sec:ctmc} and~\ref{sec:csp},
  respectively. We can use $\hat\pi_\alpha$ to estimate the free energy of
  milestone~$B_\alpha$
\begin{equation}
  \label{eq:14}
  \Deltai \hat G_\alpha = - \beta^{-1} \ln \hat\pi_\alpha
\end{equation}
where $\beta$ denotes the reciprocal of the thermodynamic temperature
of the system. 

\subsection{State-space partitioning, transition state ensemble
  identification, and other diagnostic tools}
\label{sec:diagno}

Even though the trial milestones merely serve as intermediary to
  construct the actual MSM, these milestones can still be used to
  analyze the MD data and partition the state space in ways that
  highlights important features of its dynamics.  

First we can organize the trial milestones onto a
network: if $N_{i,j}$ denotes the number of times milestone $S_j$ was
hit directly after $S_i$, we put an edge between node $i$ and node $j$
with weight
\begin{equation}
  \label{eq:15}
  d_{i,j} = \frac{N_{i,j}+N_{j,i}}{2T}
\end{equation}
where $T$ is the total length of the timeseries for $\xb(t)$. Note
that, consistent with the detailed balance condition, we have
symmetrized the weight $d_{i,j}$, i.e. the network is undirected,
$d_{i,j}=d_{j,i}$.

Next we can compute the free energy of the trial milestones. If
$T_i = \int_0^T \delta_{i_*(t),i} dt$ denotes the total time that the
last milestone hit was $S_i$ (so that $\sum_i T_i = T$, the total
length of the timeseries), we can define the probability
distribution
\begin{equation}
  \label{eq:11}
  \hat p_i = \frac{T_i}{T}
\end{equation}
The distribution $\hat p_i$ gives the proportion of time during which
$S_i$ was the last milestone hit by $\xb(t)$ (see Secs.~\ref{sec:ctmc}
and~\ref{sec:csp} for its expression in the limit as $T\to\infty$ when
the system's dynamics is governed by a Markov jump process or a
Langevin equation, respectively), and we can estimate the free energy
of the trial milestone~$S_i$ via:
\begin{equation}
  \label{eq:14b}
  \Deltai \hat A_i = - \beta^{-1} \ln \hat p_i
\end{equation}
The free energy $\Deltai \hat A_i$ defines a landscape on the network of
trial milestones.  One may expect that the nodes associated with the
target milestones will be close to the local minima of $\Deltai \hat A_i$;
however, we stress that this need not be the case, since the dynamics
on the full network (rather than the one restricted to the target
milestones alone) can be quite complicated (in particular, not
Markovian). 

To partition the state space, it is better to introduce the committor
functions of the trial milestone~$S_i$ with respect to the target
milestone $B_\alpha$, $q_\alpha(i)$. By definition, $q_\alpha(i)$
gives probability that, after hitting $S_i$, the trajectory will hit
the target milestone $B_\alpha$ before hitting any other target
milestones, and it can be estimated from the timeseries as
\begin{equation}
  \label{eq:12}
   \hat q_\alpha(i) = \frac{N_{i,\alpha}}{N_i}
\end{equation}
where $N_i = \sum_\alpha N_{i,\alpha}$ (so that
$\sum_\alpha \hat q_\alpha(i)=1$). The committor functions can be used to do a
(soft) partitioning of the network into basins of nodes that are more
likely to be attracted next to one target milestone rather than any
other: for example, those nodes~$i$ such that $q_\alpha(i)$ is close
to 1 are associated with milestones $S_i$ out of which the trajectory
is very likely to hit $B_\alpha$ next. A hard partitioning can also be
performed by assigning $i$ to
$\alpha(i) = \argmax_{\alpha\in \mathcal{M}} q_{\alpha}(i)$.

The committor probability $q_{\alpha}(i)$ whose estimator is given
in~\eqref{eq:12} can also be used to identify the transition state
ensemble (TSE), i.e. the trial milestones that lie in between the
states in the target set $\mathcal{M}$. If the target set contains
only two states, $\mathcal{M}=\{\alpha,\beta\}$, it follows that all trial
milestones $i$ satisfy $q_{\alpha}(i)=1-q_{\beta}(i)$ and the TSE is
such that $q_{\alpha}(i)\approx \frac12$. For target sets with more
than 2 states the $\frac12$ criterium may become less effective as the
TSE as it can in general connect multiple states. To get around this
difficulty, we can introduce a TSE index based on the entropy of the
committor probability $q_{\alpha}(i)$. Recalling that
$\sum_{\alpha\in \mathcal{M}}q_{\alpha}(i)=1$ for any milestone
$S_{i}$ (i.e.  $q_{\alpha}(i)$ is a probability distribution in
$\alpha$), we propose to use the normalized entropy (sometimes also
referred to as efficiency) of $q_{\alpha}(i)$ as TSE index:
\begin{equation}
  \sigma(i)=\begin{cases}
    -\sum_{\alpha\in \mathcal{M}}q_{\alpha}(i)\ln
    q_{\alpha}(i)/\ln n(i) 
    & n(i)>1\\
    0 & n(i)=1
  \end{cases}\label{eq:TSE}
  \end{equation}
were the sum is carried out over the non-zero entries of $q_\alpha(i)$
and $n(i)$ is the number of such entries (more generally, we could
restrict the sum to the entries of $q_\alpha(i)$ that are above some
small threshold $\delta$). The TSE index $\sigma(i)$ is
1 if the non zero values of $q_{\alpha}(i)$ are all identical, and it
is 0 if only one value of $q_{\alpha}(i)$ is different than
zero. Therefore, the closer $\sigma(i)$ is to 1, the higher the chance
that state $i$ be a member of the transition state ensemble. Once the
states in the TSE have been identified, we can go back to their
committor values to determine between which target states they lay.

To characterize the physical origin of the
metastability of the target milestones, it is also useful to decompose
their free energy into an energetic component and an
entropic one. This can be done as follows.  Given the system's
potential energy $U(\xb)$, a mean energy can be assigned to each of
the target milestones via
\begin{equation}
  \label{eq:mileEi}
  \Deltai \hat E_{\alpha}=\frac1{T_\alpha}\int_{0}^{T}
  U(\xb(t))\delta_{\alpha_{*}(t),\alpha}dt - \bar E
\end{equation}
where $T_\alpha$ the total time the last target milestone hit by $\xb(t)$ is
$B_\alpha$ and
\begin{equation}
  \label{eq:barE}
  \bar{E}=\frac{1}{T}\int_{0}^{T}U(\xb(t))dt 
\end{equation}
so that $\sum_{\alpha} \Deltai \hat E_{\alpha} \hat \pi_\alpha = 0$.
We can then estimate the entropy $\Deltai \hat S_\alpha$ of the target 
milestone via 
\begin{equation}
  \label{eq:TdS}
  (k_B \beta)^{-1} \Deltai \hat S_\alpha=\Deltai \hat E_\alpha-\Deltai
  \hat G_\alpha
\end{equation}
where $\Deltai \hat G_\alpha$ is the free energy estimated in~\eqref{eq:14}
and $k_B$ is Boltzmann constant. Target milestones with comparable
$\Deltai G_\alpha$ have similar statistical weights, and the lower
$\Deltai G_\alpha$ the more thermodynamically stable they are (meaning
the timeseries $\xb(t)$ tends to return to them more often,
comparatively): by comparing their values of $\Deltai \hat E_\alpha$ and
$\Deltai \hat S_\alpha$ we can then determine whether their stability is of
energetic or entropic origin, respectively.

The usefulness of the diagnostic tools introduced above will be illustrated in
the examples treated in Secs.~\ref{sec:glyalagly} and~\ref{sec:Beta3s}.

\section{Theoretical justification}
\label{sec:theo}

Let us now justify the use of the metastability index defined
in~\eqref{eq:6} to identify target milestones over which the dynamics
can be mapped in an approximately Markovian way.

\subsection{The case of Markov chains}
\label{sec:ctmc}

We begin by discussing the simpler case when the dynamics of $i_*(t)$
on the trial milestones is itself Markovian -- as we mentioned in
Sec.~\ref{sec:algo}, we do not make this assumption in general, but it
is a convenient starting point for our theoretical explanation. The
general situation when $i_*(t)$ is not Markovian will be discussed in
Sec.~\ref{sec:csp}.

If $i_*(t)$ is Markov, its dynamics is specified by a rate matrix $L$
whose entries we will denote by $L_{i,j}$ to distinguish them from the
rate matrix entries $k_{\alpha,\beta}$ on the target
milestones. Assuming detailed balance, $L_{i,j}$ satisfies
\begin{equation}
  \label{eq:16}
  p_i L_{i,j} = p_j L_{j,i}, \qquad i,j = 1,\ldots, N
\end{equation}
where $N$ is the number of trial milestones and $p_i$ is their
equilibrium probability density -- the estimator for $p_i$ was
given in~\eqref{eq:11}. Together with the assumption of ergodicity
(i.e. that time averages along $i_*(t)$ converge towards ensemble
averages over $p_i$), \eqref{eq:16} implies that all the $N$
eigenvalues of $L_{i,j}$ are real, with one being 0 and all the other
negative. We will denote these eigenvalues by $\lambda_i$, and order
them so that $0=\lambda_0 \le |\lambda_1| \le \cdots \le
|\lambda_{N-1}|$.

The eigenvalues of $L$ permit to give a precise definition of
what it means for the chain to display metastability. In turns this
indicates how to coarse-grain this chain, which in our context means
how to chose good target milestones. Specifically, a chain will be
metastable if its eigenvalues can be separated into two well separated
groups, i.e. if there exists an $M<N$ such that
\begin{equation}
  \label{eq:17}
  \lambda_{M-1}/\lambda_M \ll 1
\end{equation}
If such a separation exists, it means that the $M$ eigenvalues with
index smaller than $M$ describe relaxation processes in the chain that
occur on timescales that are much slower than those described by the
$N-M$ eigenvalues with index larger or equal to $M$. In turn, this
implies that these slow processes can be approximately described by a
smaller chain with only $M$ states. The practical questions then
become: (i) how to assess whether \eqref{eq:17} is satisfied for some
$M$ without having to compute the full spectrum of $L$ (since
this computation is hard in general), and (ii) how to reduce the
dynamics to a chain with only $M$ states?

In Ref.~[\onlinecite{Bovier:2002ab}], Bovier and collaborators
addressed these two questions, and the answers they provided are the
basis for the algorithm we presented in Sec.~\ref{sec:algo}. First
they proved that \eqref{eq:17} holds if and only if an indicator
closely related to the metastability index $\rho_{\mathcal{M}}$
in~\eqref{eq:6} is small for some
$\mathcal{M}=\{\alpha_1, \alpha_2,\ldots, \alpha_M\}$. If that is the
case, the ratio in~\eqref{eq:17} is in fact proportional to
$\rho_{\mathcal{M}}^2$. Second, they showed how to reduce the chain
onto a smaller chain involving only the nodes identified by
$\mathcal{M}$ (i.e. in our context, involving only the target
milestones) in such a way that the $M$ eigenvalues of this reduced
chain be close to the $M$ first eigenvalues of the original chain
(where closeness can again be measured in terms of
$\rho_{\mathcal{M}}$).

It should be stressed that \eqref{eq:17} can be satisfied with more
than one value of $M$. This simply means that there can be more than
one low-lying group of eigenvalues. In turns this implies that there
can be more than one choice of good target milestones.

For completeness, let us end this section by giving explicit
  expressions for some of the quantities that were defined in
  Sec.~\ref{sec:algo} in the context of a system whose dynamics is
  governed by an MJP with generator $L$. First, the probability
  $\Gamma_{\!i,j}$ whose estimator was given in \eqref{eq:2} can be
  expressed as:
\begin{equation}
  \label{eq:GammaComm}
  \Gamma_{\! i,j} =\sum_{k=1}^N P_{i,k}q_{i,j}(k)
\end{equation}
Here $P_{i,j}$ (not to be confused with the entries of the transfer
  operator $T(\tau)= e^{L\tau}$, $\tau>0$) are the entries of the transition matrix
  defined as
\begin{equation}
  \begin{aligned}
    P_{i, j} = \frac{L_{i, j}}{\sum_{j \neq i} L_{i, j}}\quad
    (i\not=j), 
  \end{aligned}
  \label{eq:TransMat}
\end{equation}
which gives the probability that the state first visited by the chain
after $i$ is $j\not=i$; and $q_{i,j}(k)$ is the committor probability
solution of
\begin{equation}
  \begin{cases}
    \sum_{l=1}^N L_{k,l} q_{i,j}(l) = 0, \qquad&
    k\notin\{i,j\}\\
    q_{i,j} (k) = 0, & k=i \\
    q_{i,j} (k) = 1, & k=j
  \end{cases} 
  \label{eq:commLinSys}
\end{equation}
The probabilities $\Gamma_{\!i,j}$ can also be conveniently calculated
in terms of mean recurrence times (MRT) and mean first passage times
(MFPT) via the formula
\begin{equation}
\Gamma_{\!i,j}=\frac{\tau_{i}}{\tau_{i,j}+\tau_{j,i}}
\label{eq:GammaMFPT}
\end{equation}   
where $\tau_{i}$ is the MRT of the state $i$ and $\tau_{i,j}$ is the
MFPT to go from state $i$ to state $j$. A proof of  relation
\eqref{eq:GammaMFPT} can be found in the Appendix of this paper. In
addition, the committor probability $q_{\alpha}(i)$ whose estimator
was given in \eqref{eq:12} solves an equation similar
to~\eqref{eq:commLinSys}, with different boundary conditions:
\begin{equation}
  \begin{cases}
    \sum_{j=1}^N L_{i,j} q_{\alpha}(j) = 0, \qquad&
    i\notin\mathcal{M}\\
    q_{\alpha} (i) = 0, & i \in \mathcal{M}\setminus \{\alpha\} \\
    q_{\alpha} (i) = 1, & i=\alpha
  \end{cases} 
  \label{eq:commLinSys2}
\end{equation}
We can relate the distributions $\pi_\alpha$ and $p_i$, whose
estimator were given in~\eqref{eq:13} and~\eqref{eq:11} respectively, as
 \begin{equation}
   \label{eq:19}
   \pi_\alpha = \sum_{i=1}^N p_i q_\alpha(i)
 \end{equation}
 which also means that the corresponding free energies whose
 estimators were given in~\eqref{eq:14} and~\eqref{eq:14b}
 respectively, are related as
 \begin{equation}
   \label{eq:23}
   e^{-\beta \Deltai G_\alpha} = \sum_{i=1}^N e^{-\beta \Deltai A_i} q_\alpha(i)
\end{equation}
Finally, the rate matrix entries on the target milestone,
$k_{\alpha,\beta}$, whose estimator was given in~\eqref{eq:5b} can be
expressed as
\begin{equation}
  \label{eq:22}
  k_{\alpha,\beta} = \frac1{\pi_\alpha} \sum_{i=1}^N p_i q_\alpha(i)
  L_{i,\beta} \qquad (\alpha \not = \beta)
\end{equation}
These formulas can be justified within the framework of transition
path theory (TPT)
\cite{weinan2006towards,Metzner:2009ab,vanden2010transition} and are
useful for analysis. However we stress that we do not need to solve
any of the equations above to apply the procedure outlined in
Sec.~\ref{sec:algo}. Indeed, this procedure can be used with the
estimators given in that section, which only require the
timeseries~$\xb(t)$ as input.

\subsection{Generalization to continuous state-spaces}
\label{sec:csp}

In Refs.~[\onlinecite{Bovier:2004aa},\onlinecite{Bovier:2005aa}], the
results of Ref.~[\onlinecite{Bovier:2002ab}] were generalized to
situations were the Markovian dynamics takes place on a continuous
state-space, like $\xb(t)$ does (or more generally the pair
$(\xb(t), \pb(t))$ if the momentum $\pb(t)$ needs to be added to make
the description Markovian, as in~\eqref{eq:15} below). The main
technical difficulty in that case is that the metastability index
$\rho_{\mathcal{M}}$ defined in~\eqref{eq:6} measures the probability
to go to a state after leaving another one, rather than returning to
that state. The problem is that, in the continuous state-space
setting, these states cannot be identified with points in the state
space, since the probability to hit a point is zero as soon as the
space dimension is higher than 1.

To get around this difficulty, it was shown in
Refs.~[\onlinecite{Bovier:2004aa},\onlinecite{Bovier:2005aa}] that one
can redefine states by fattening any specific point into a little
domain that contains it, so that a metastability index
$\rho_{\mathcal{M}}$ can again be defined as in~\eqref{eq:6} and
allows to identify low-lying groups of eigenvalues when they exist.
This fattening procedure is similar to that of defining trial
milestones: they are indeed regions containing specific points in the
original state-space of the system, which provides a theoretical
justification to the algorithm proposed in Sec.~\ref{sec:theo}. It is
important to note, however, that
Refs.~[\onlinecite{Bovier:2004aa},\onlinecite{Bovier:2005aa}] only
gave prescriptions on how to perform this fattening (i.e. how to define
trial milestones) in very specific (and simple) cases, like that of an
system governed by overdamped dynamics in the limit of very small
temperature. We are obviously interested in more complicated
situations here, in which case it is no \textit{a~priori} obvious how
to define the trial milestones. A few procedures to do so will be
discussed in Secs.~\ref{sec:glyalagly} and~\ref{sec:Beta3s}: these
procedures are by no means the only ones one could envision, but they
proved sufficient in these examples and should be transportable to
other ones. Note that this also means that one should verify
  \textit{a~posteriori} that the dynamics on the target milestones
  identified by the procedure is indeed approximately Markovian. As
  usual, this can be done by checking that the first passage time
  between target milestones adjacent on the network of the MSM are
  exponentially distributed. In the examples we treated below, this
  turned out to be the case, indicating that the MSMs we constructed
  were indeed accurate. 

Assuming that we have picked trial milestones and identified
  target ones, for completeness let us give explicit formulas for some
  of the quantities introduced in Sec.~\ref{sec:algo} in the context
  of a system whose dynamics is governed by the Langevin equation
  \begin{equation}
    \label{eq:15}
    \begin{cases}
      \dot \xb = m^{-1}\pb\\
      \dot \pb = - \nabla U(\xb) -\gamma\pb
      + \sqrt{2\beta^{-1}}\, m^{1/2} \gamma^{1/2}\, \etab(t),
    \end{cases}
\end{equation}
where $U(\xb)$ is the potential energy of the system, $m$ the mass
matrix, $\gamma$ the friction tensor, and $\etab(t)$ is a white-noise
satisfying $\< \etab(t)\> = 0$,
$\<\etab(t) \etab^T(s) \> = \text{Id}\, \delta(t-s)$ -- other choices
of thermostats are possible, and the formula below can be adapted to
those straightforwardly. System~\eqref{eq:15} is ergodic with respect
to Boltzmann-Gibbs probability density function
\begin{equation}
  \label{eq:21}
  \rho(\xb,\pb) = Z^{-1} e^{-\beta H(\xb,\pb)}
\end{equation}
where $H(\xb,\pb) = \frac12 \pb^T m^{-1} \pb + U(\xb)$ is the
Hamiltonian and
$Z = \int_{\Omega \times \RR^{3n}} e^{-\beta H(\xb,\pb)} d\xb d\pb$.
The expression for the probability $\Gamma_{\!i,j}$ whose estimator was
given in~\eqref{eq:2} is quite complicated if, unlike what we did in
Sec.~\ref{sec:ctmc}, we do not assume that the dynamics on the trial
milestones is Markovian.  We can, however, give explicit expression
for the rate matrix entries $k_{\alpha,\beta}$, the distribution $\pi_\alpha$,
and the free energy $G_\alpha$ whose estimators were given
in \eqref{eq:5b}, \eqref{eq:13}, and~\eqref{eq:14}, respectively. These
expressions were derived in Ref.~\cite{Schutte:2011aa} and they
involve the committor function~ $Q_\alpha \equiv Q_\alpha(\xb,\pb)$
solution of
\begin{equation}
  \label{eq:20}
  \begin{aligned}
    0 &= m^{-1} \pb \cdot \nabla_{\!\xb} Q_\alpha - \nabla U(\xb)
    \cdot \nabla_{\!\pb} Q_\alpha\\
    & \quad - \gamma\pb \cdot \nabla_{\!\pb} Q_\alpha + \beta^{-1}
    m\gamma: \nabla_{\!\pb} \nabla_{\!\pb} Q_\alpha
  \end{aligned}
\end{equation}
with the boundary condition $Q_\alpha(\xb,\pb)=1$ if
$\xb \in \partial S_\alpha$ and $\hat \nb_\alpha(\xb) \cdot \pb >0$
and $Q_\alpha(\xb,\pb)=0$ if
$\xb \in \cup_{\beta \in \mathcal{M}\setminus \alpha } \partial
S_\beta$
and $\hat \nb_\beta(\xb) \cdot \pb >0$, where $\hat \nb_\alpha(\xb)$
denotes the unit normal vector pointing outward $\partial S_\alpha$ at
point~$\xb\in\partial S_\alpha$ and similarly for
$\hat \nb_\beta(\xb)$.  The committor function $Q_\alpha(\xb,\pb)$
gives the probability that the trajectory initiated at $(\xb,\pb)$
reaches $S_\alpha$ before any $S_\beta$ with $\beta\not=\alpha$; using
the invariance of the dynamics under $t\to-t$ and $\pb\to-\pb$,
$Q_\alpha(\xb,-\pb)$ also gives the probability that the trajectory
arriving at $(\xb,\pb)$ was last in $S_\alpha$ rather than in any
$S_\beta$ with $\beta\not=\alpha$. We then have
\begin{equation}
  \label{eq:11a}
  \pi_\alpha = \int_{\Omega\times\RR^{3n}} \rho(\xb,\pb)
  Q_\alpha(\xb,-\pb) d\xb d\pb
\end{equation}
so that $\Deltai G_\alpha = \beta^{-1} \ln \pi_\alpha$. Also
\begin{equation}
  \label{eq:25}
  \begin{aligned}
    k_{\alpha,\beta} &= \pi_\alpha^{-1}\int_{\partial S_\beta \times \RR^{3n}}
    \rho(\xb,\pb) Q_\alpha(\xb,-\pb) Q_\beta(\xb,\pb)\\
    & \qquad \qquad \times |\nb_\beta(\xb)\cdot m^{-1}
    \pb|\,  d\sigma_{\beta}(\xb) d\pb
  \end{aligned}
\end{equation}
where $d\sigma_{\beta}(\xb)$ denotes the surface element on
$\partial S_\beta$. These formulas can again be derived from
TPT~\cite{weinan2006towards,vanden2010transition}. Similar expressions
can be given for $p_i$ and $\Deltai A_i$, whose estimators were
given in~\eqref{eq:11} and~\eqref{eq:14b} by modifying the boundary
conditions in~\eqref{eq:20}.  

\begin{figure}[t]
\begin{centering}
\includegraphics[scale=0.42]{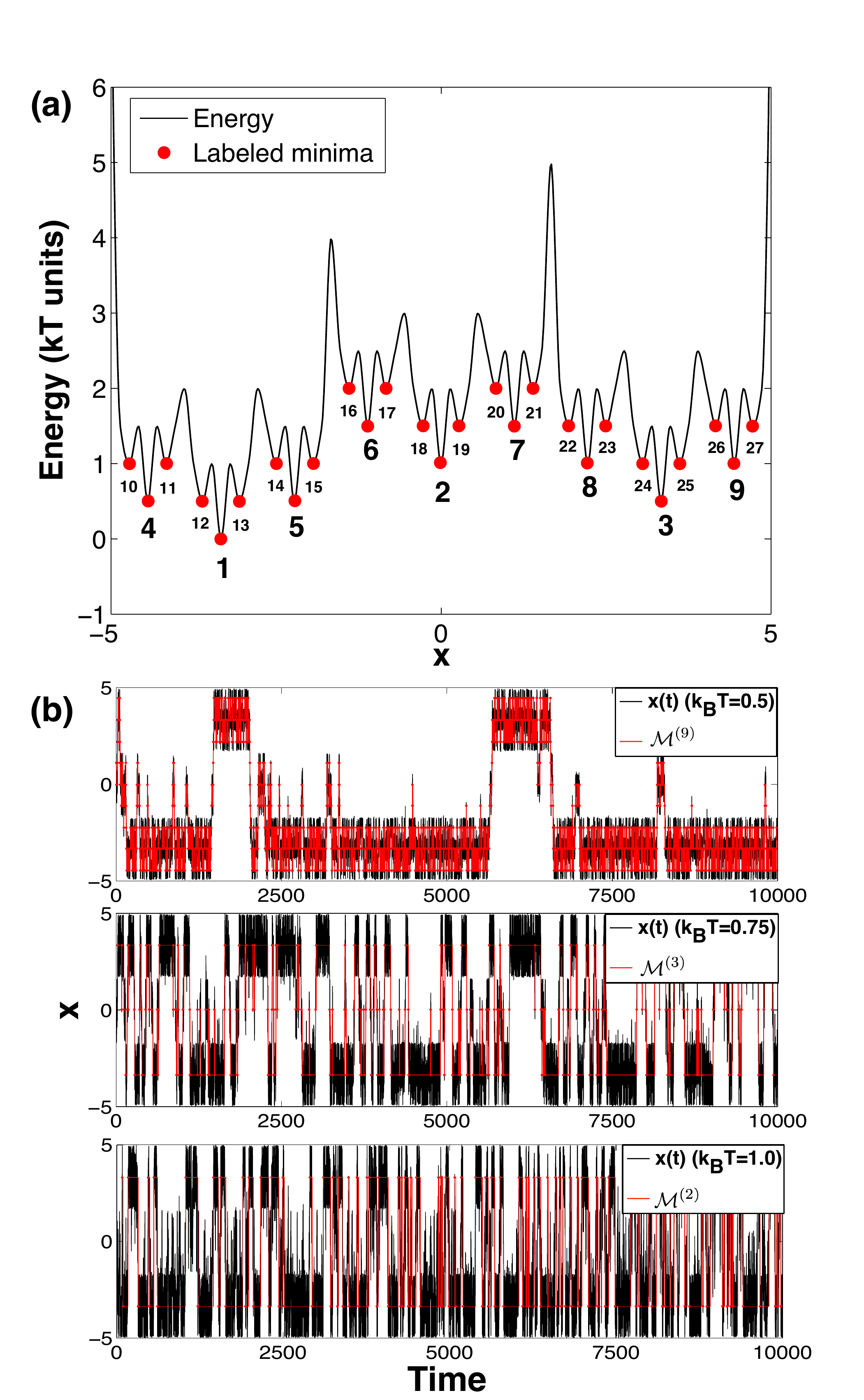}
\end{centering}
\caption{\label{fig:1Dland} (a) The one-dimensional multiple well energy
  landscape example. The red dots identify and label the milestones
  contained in the sets of target milestones. Their corresponding
  index sets are $\mathcal{M}^{(2)}=\{1,2\}$,
  $\mathcal{M}^{(3)}=\{1,2,3\}$, $\mathcal{M}^{(9)}=\{1,2,...,9\}$ and
  $\mathcal{M}^{(27)}=\{1,2,...,27\}$. (b) Overdamped Langevin
  trajectories solution of~\eqref{eq:overDump} at three temperatures
  $k_BT =0.5$, 0.75 and 1.0 (black curves). The index function $\alpha_*(t)$
  associated with the target sets $\mathcal{M}^{(9)}$,
  $\mathcal{M}^{(2)}$ and $\mathcal{M}^{(3)}$ are superimposed on the
  trajectories $x(t)$ (red curves).}
\end{figure}

\subsection{Illustrative example}
\label{sec:illustrativeex}

It is useful to illustrate the results of this section on a simple
example. Specifically, we consider the motion of a particle by
overdamped Langevin dynamics on the one-dimensional potential energy depicted
in Fig.~\ref{fig:1Dland}(a). The governing equation is
\begin{equation}
  \gamma\dot{x}(t)=-U'(x)+\sqrt{2\beta^{-1}\gamma}\, \eta(t)\label{eq:overDump}
\end{equation}
where $U'(x)$ denotes the derivative of potential energy $U(x)$,
$\gamma$ is the friction coefficient (which we will set to $\gamma=1$)
and $\eta(t)$ a white-noise such that $\langle\eta(t)\rangle=0$ and
$\langle\eta(t)\eta(t')\rangle=\delta(t-t')$. Typical trajectories
solution of~\eqref{eq:overDump} at three different temperatures are shown
in Fig.~\ref{fig:1Dland}(b).

\begin{figure}[t]
\begin{center}
\includegraphics[scale=0.4]{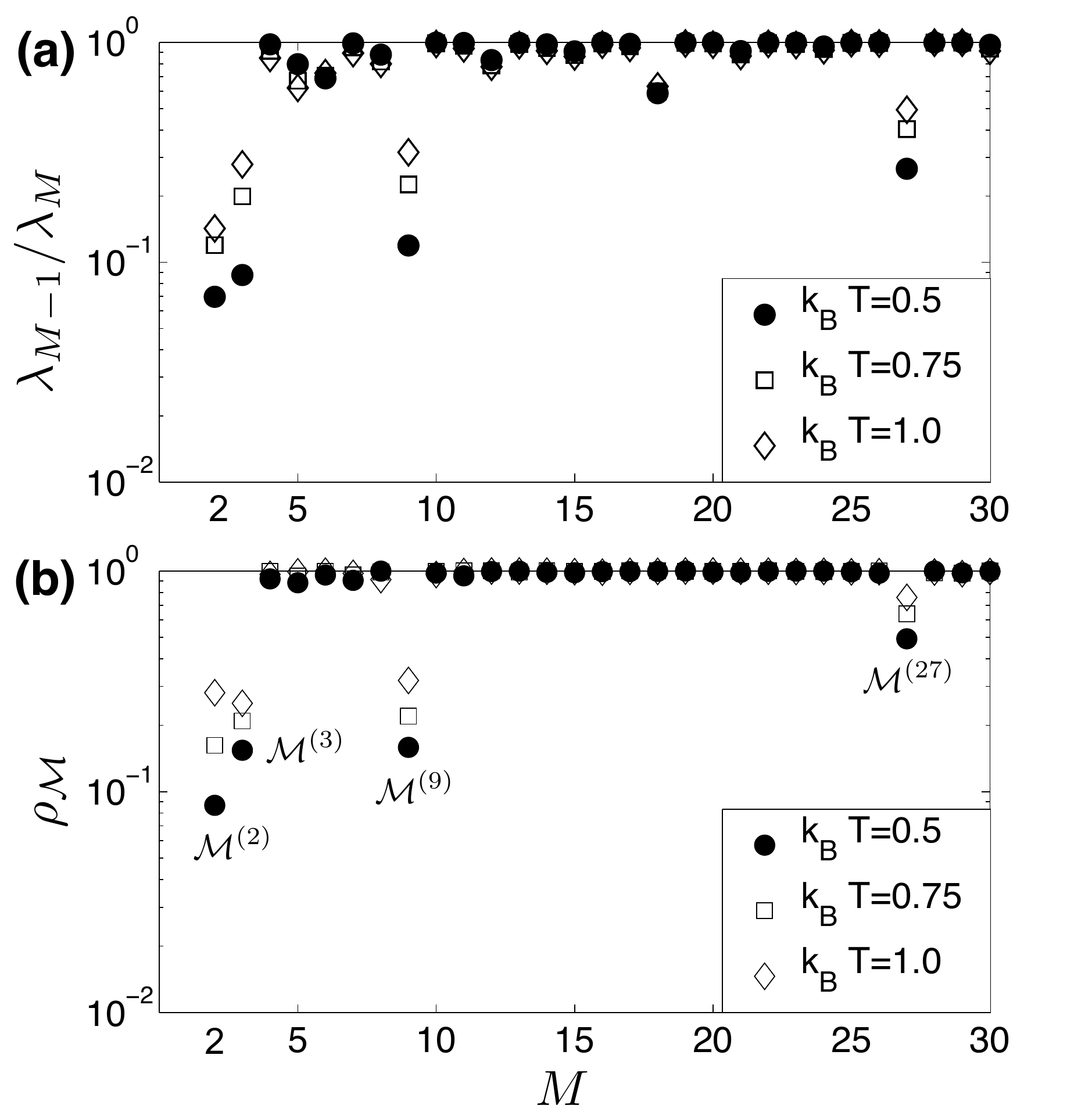}
\end{center}
\caption{\label{fig:1DrhoB}(a) First 30 ratios of consecutive
  eigenvalues of the generator~\eqref{eq:18} in a range of
  temperatures from 0.5 to 1.0. The small ratios are
  $\lambda_{1}/\lambda_{2}$, $\lambda_{2}/\lambda_{3}$,
  $\lambda_{8}/\lambda_{9}$ and $\lambda_{26}/\lambda_{27}$.  (b) The
  values of the metastability index $\rho_{\mathcal{M}}$ minimized
  over index sets $\mathcal{M}$ of increasing cardinal $M$:
  $\rho_{\mathcal{M}}$ clearly correlates well with the eigenvalue
  ratios shown in panel (a), and the associated index sets
  $\mathcal{M}$ permit to identify the target milestones shown in
  Fig.~\ref{fig:1Dland} that capture the slow processes in the
  system.}
\end{figure}

As can been seen in Fig.~\ref{fig:1Dland}(a), the potential has a
hierarchical structure with a total of 27 local minima separated by
barriers of various heights. Correspondingly, the generator of the
process governed by~\eqref{eq:overDump}, namely the operator
\begin{equation}
  \label{eq:18}
  \mathcal{L} = -U'(x) \partial_x + \beta^{-1}\, \partial^2_x
\end{equation}
has a spectrum with several groups of low-lying eigenvalues. This
spectrum was obtained by spatial-discretization of~\eqref{eq:18} on a
grid of 200 points, and the ratio of successive eigenvalues are shown
in Fig.~\ref{fig:1DrhoB}(a) at several different temperatures. Small
ratios identify low-lying groups of eigenvalues, and several of them
can be seen: $\lambda_1/\lambda_2$, $\lambda_2/\lambda_3$,
$\lambda_8/\lambda_9$, and finally $\lambda_{26}/\lambda_{27}$ are all
small at temperatures ranging from $\beta^{-1} =0.5$ to $\beta^{-1}=1.0$. The
eigenvalues involved in these groups describe processes arising on
slow time scales that can be organized as follows: $\lambda_1$ is
associated with longest timescale of hopping over the largest barrier
separating the left basin left centered around milestone 1 and right
one centered around milestone 2 in Fig.~\ref{fig:1Dland}(a);
$\lambda_2$ is associated with the next longest time hopping over the
barrier separating the basin at the extreme left centered around
milestone 1 and the one in the center centered around milestone 3;
$\lambda_3$, ..., $\lambda_8$ are associated with hoping over the
barriers separating milestones $1$, $4$, $5$, milestones $3$, $6$, $7$
and milestones $2$, $8$, $9$; and finally $\lambda_9$, ...,
$\lambda_{27}$ are associated with hoping over the barriers separating
milestones $4$, $10$, $11$, milestones $1$, $12$, $13$, etc. From the
results in Secs.~\ref{sec:ctmc} and~\ref{sec:csp}, the existence of
the low lying groups also suggests that the continuous-time dynamics
can be approximated by Markov chains (i.e MSMs) containing,
respectively, 2, 3, 9 and 27 states.

\begin{figure}[t]
\includegraphics[scale=0.45]{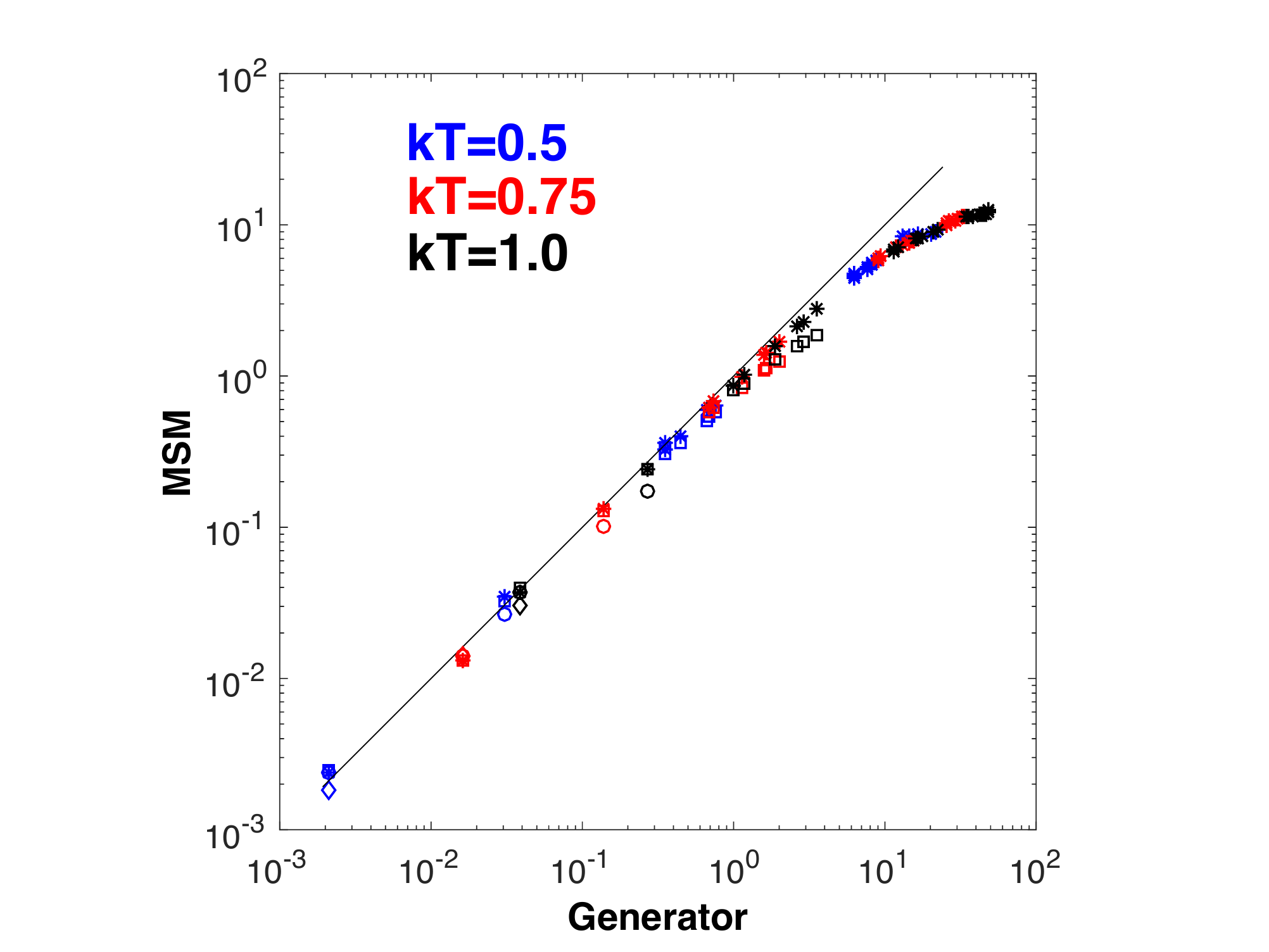}
\caption{\label{fig:crossval} The eigenvalues of the MSMs made of the
  states in $\mathcal{M}^{(2)}$ (circles), $\mathcal{M}^{(3)}$
  (diamonds), $\mathcal{M}^{(9)}$ (squares), and $\mathcal{M}^{(27)}$
  (stars) are plotted against the eigenvalues of the generator of the
  original process with the same indices. As can be seen, the MSMs
  capture the low-lying part of the spectrum accurately, with
  deviations observed only for the largest eigenvalues in the
  low-lying group, where metastability becomes weaker. }
\end{figure}

To confirm this prediction, we used the algorithm presented in
Sec.~\ref{sec:algo} to construct these MSMs. Specifically, we used the
200 discretization points uniformly spaced between $x=-5$ and $x=5$ as
trial milestones and computed the matrix entries $\Gamma_{\! i,j}$
defined in~\eqref{eq:2} from a trajectory obtained by
integrating~\eqref{eq:overDump}. We then minimized the metastability
index $\rho_{\mathcal{M}}$ over index sets $\mathcal{M}$ containing
from $M=2$ up to $M=30$ indices. The minimum values of
$\rho_{\mathcal{M}}$ for each value of $M$ are shown in
Fig.~\ref{fig:1DrhoB}(b) and, consistent with Bovier's result, they
correlate well with the values of the eigenvalue ratios
$\lambda_{M-1}/\lambda_M$ shown in Fig.~\ref{fig:1DrhoB}(a), thereby
confirming that we can use the metastability index
$\rho_{\mathcal{M}}$ to identify low-lying groups of eigenvalues. In
particular, $\rho_{\mathcal{M}}$ could be made small when $M=2$, 3, 9
and 27, and the corresponding index sets were
$\mathcal{M}^{(2)}=\{1,2\}$, $\mathcal{M}^{(3)}=\{1,2,3\}$,
$\mathcal{M}^{(9)}=\{1,...,9\}$, and
$\mathcal{M}^{(27)}=\{1,...,27\}$. The associated target milestones
are shown in Fig.~\ref{fig:1Dland}(a). They clearly identify the
lowest point of the wells on the hierarchical potential landscape
$U(x)$. The trajectory projected onto these target milestones when
$\mathcal{M}^{(2)}=\{1,2\}$, $\mathcal{M}^{(3)}=\{1,2,3\}$, and
$\mathcal{M}^{(9)}=\{1,...,9\}$ are shown in red in
Fig.~\ref{fig:1Dland}(b): these red pieces can be used to calculate
the rate matrices $k_{\alpha,\beta}$ of the different MSMs on
$\mathcal{M}^{(2)}$, $\mathcal{M}^{(3)}$, $\mathcal{M}^{(9)}$, and
$\mathcal{M}^{(27)}$ via maximum likelihood maximization, as explained
in Sec.~\ref{sec:MSM}. We can then calculate the spectrum of these
rate matrices and compared them to the spectrum of the
generator~\eqref{eq:18} of the original process. The result of these
calculation is shown in Fig.~\ref{fig:crossval} which shows that the
eigenvalues of these different MSMs do indeed match the low-lying ones
of the generator of the original process.

Note that the calculations above to identify target milestones were
conducted at different temperatures, and showed that the target
milestones were robust for temperature ranging from $\beta^{-1}= 0.5$ to
$\beta^{-1}=1.0$, even though their corresponding metastability index
$\rho_{\mathcal{M}}$ slowly grew with temperature. This is consistent
with the fact that the ratios $\lambda_{M-1}/\lambda_M$ also grow with
temperature, as the system becomes slowly less metastable as
it is heated up, with the groups associated with the lowest barriers
disappearing first.

Note also that in these calculations, few transitions events between
the target milestones were observed (much less in particular that what
is required to calculate the rate matrix entries $k_{\alpha,\beta}$
accurately), and yet the procedure was able to identify these
milestones correctly.

\begin{figure}[t]
\includegraphics[scale=0.45]{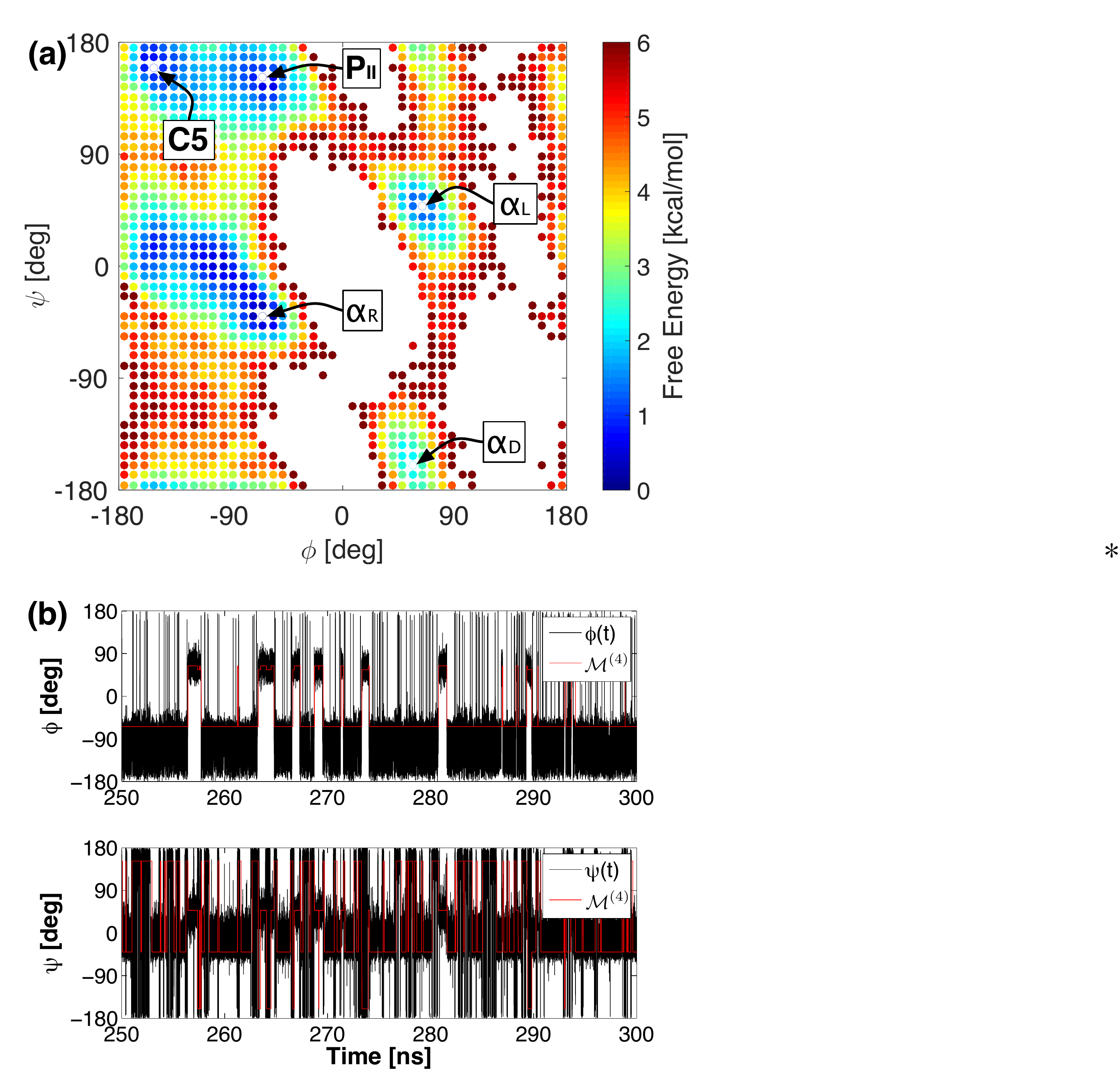}
\caption{\label{fig:GAGland} (a) Trial milestones used to analyze
    the trajectory from the MD simulation of the GAG peptide at
    $T=330$ K. These trial milestones are disks centered around grid
    points in the $(\phi,\psi)$-space, and they are colored according
    to their free energy $\Deltai A_i$, whose estimator is given
    in~\eqref{eq:14b}. The states composing the target milesones
  $\alpha_{R}$, $\alpha_{L}$, $P_{\text{II}}$ and $\alpha_{D}$ are
  represented as white filled circles. (b) A portion of a $1.3$ $\mu$s
  MD simulation at $T=330$ K projected onto $\phi$ and $\psi$ (black
  curve). In red we show the index function $\alpha_*(t)$ of the
  states in the target set
  $\mathcal{M}^{(4)}=\{\alpha_R,\alpha_L, P_{\text{II}},\alpha_D\}$.}
\end{figure}

\begin{figure}[t]
\includegraphics[scale=0.55]{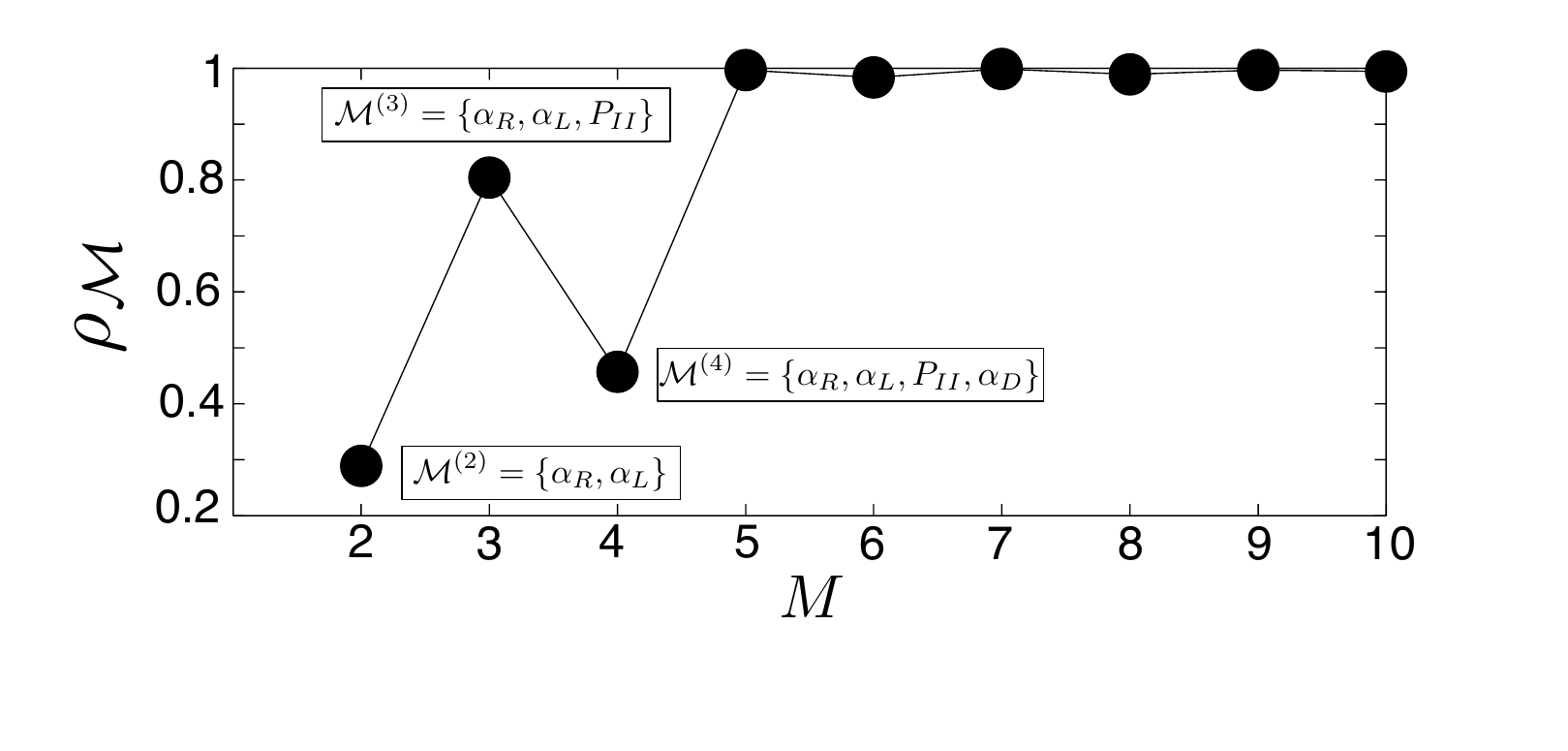}
\caption{\label{fig:GAGrhoB} GAG peptide: Minimum values of the metastability index
  $\rho_{\mathcal{M}}$ as a function of $M$, along with the index sets
  $\mathcal{M}$ associated with $\rho_{\mathcal{M}}$ smaller than 1. A
  Monte Carlo scheme was utilized to perform the minimization.}
\end{figure}

\begin{figure*}[t]
\includegraphics[clip,scale=0.4]{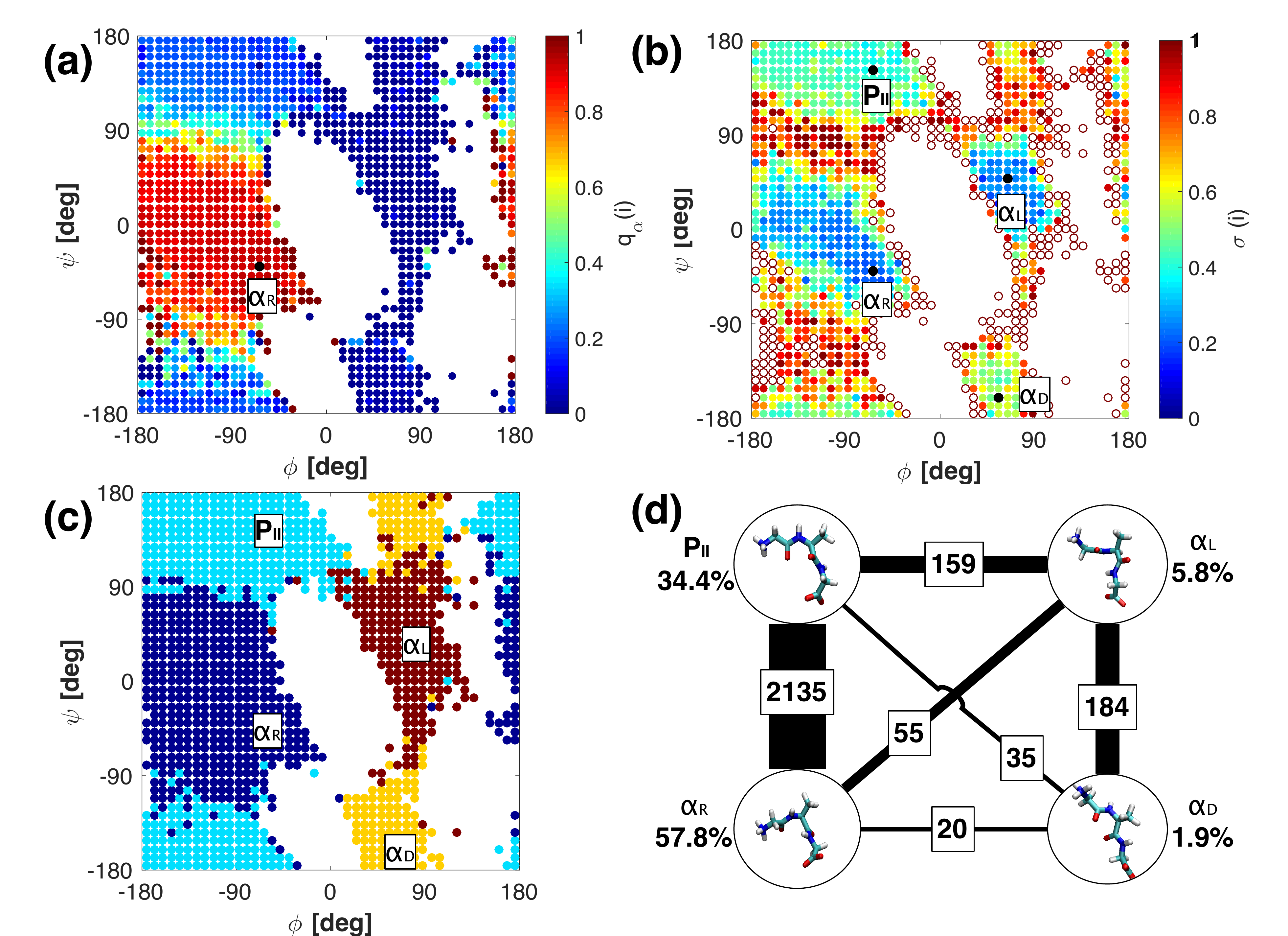}
\caption{\label{fig:GAG-commTScluster} GAG peptide: Analysis based on the MSM with
  states in $\mathcal{M}^{(4)}$. (a) Committor probabilities
  $q_{\alpha_{R}}(i)$. (b) The transition state index $\sigma(i)$, the
  red regions represents the milestones likely to be part of a TSE;
  unfilled circles are trial milestones that were visited only once
  (and hence for which $q_\alpha(i)=1$ for some $\alpha$). (c) Hard
  partitioning of the trial milestones using
  $\alpha(i) =\argmax_{\alpha\in\mathcal{M}}q_{\alpha}(i)$. (d)
  Network of the MSM : The numbers of transitions observed between
  pairs are shown on the edges, and the equilibrium probability of the
  target milestones are shown next to the nodes.}
\end{figure*}

\section{Gly-Ala-Gly peptide}
\label{sec:glyalagly}

In this section, we use the method outlined in Sec.~\ref{sec:algo} to
analyze a MD trajectory of a solvated Glycine-Alanine-Glycine peptide
(GAG). The GAG peptide was modeled using the CHARMM 27 force field and
simulated in a box of 475 TIP3P water molecules using the program NAMD
version 2.8 \cite{Phillips:2005fk}.  After minimization the system was
equilibrated for 10ns with the peptide held constrained. The
equilibration was followed by a 1.3 $\mu$s production run with
Langevin dynamics at 330 K, using a friction constant of 5 ps. The
bonds between hydrogens and heavy atoms were kept rigid to allow
integration at 2 fs. Frames were saved every 0.5 ps and a total of
$2.6\cdot10^{6}$ trajectory snapshots were collected.

To construct the set of trial milestones, we projected the MD
trajectory onto the pair of dihedral angles $(\phi,\psi)$ that
corresponds to the central residue alanine. We discretized the
$(\phi,\psi)$-space into a square grid of size $8$ deg and used
circles of radius $4$ deg around these discretization points as trial
milestones -- we also used a finer grained definition of the trial
milestones, a grid of points at 5 deg but this did not significantly
change the results. Only those milestones that we were hit by the
trajectory were kept: these consist of the $N=1303$ milestones shown
as circles in Fig.~\ref{fig:GAGland}(a), and colored according to
  their free energy $\Deltai A_i$ whose estimator is given
  in~\eqref{eq:14b}.  This free energy defines a landscape that is
  divided in four macro-regions, often labeled as C5, $P_{\text{II}}$,
  $\alpha_R$, $\alpha_L$ and $\alpha_D$. Interestingly, the overall
  topography of this landscape is consistent with that obtained in a
  reference study on a solvated alanine dipeptide \cite{Smith:1999aa}
  where the $\alpha_{R}$ conformation is the most populated (see the
  Ramachandran map in Ref.~[\onlinecite{Iwaoka:2002aa}] for the
  conformation names and
  Refs.~[\onlinecite{Iwaoka:2002aa,Ding:2003aa,Weise:2003aa}] for
  comparison with NMR studies on the alanine conformational
  preferences).  Since the target milestones that emerged from our
  analysis were located in the regions around C5, $P_{\text{II}}$,
  $\alpha_R$, $\alpha_L$ and $\alpha_D$ (these target milestones are
  shown as filled white circles in Fig.~\ref{fig:GAGland}(a)), we used
  this nomenclature to designate them.  Fig.~\ref{fig:GAGland}(b)
  shows a portion of the trajectory projected on the angles $\phi$ and
  $\psi$.

From the MD trajectory the matrix of probabilities 
$\Gamma_{\! i,j}$  was estimated from~\eqref{eq:2}, and  
the metastability index $\rho_{\mathcal{M}}$ was minimized over index sets
$\mathcal{M}$ of different cardinals $M$. The results are shown in
Fig.~\ref{fig:GAGrhoB}. Two sets of target milestones were found to be
clearly metastable: $\mathcal{M}^{(2)}=\{\alpha_R,\alpha_L\}$ with
metastability index $\rho_{\mathcal{M}}=0.29$, and
$\mathcal{M}^{(4)}=\{\alpha_R,\alpha_L, P_{\text{II}}, \alpha_D\}$ with
metastability index $\rho_{\mathcal{M}}=0.45$. A third set of target
milestones, $\mathcal{M}^{(3)} = \{\alpha_R,\alpha_L,P_{\text{II}}\}$
was also metastable, but with a higher metastability index
$\rho_{\mathcal{M}}=0.8$. Interestingly, the index set minimizing
$\rho_{\mathcal{M}}$ when $M=5$ was $\mathcal{M}^{(5)} =
\{\alpha_R,\alpha_L,P_{\text{II}},\alpha_{D},C5\}$, but its metastability index
$\rho_{\mathcal{M}}$ was close to 1, i.e. it was not deemed suitable
by our analysis to be used to construct an MSM.

\begin{figure}[t]
\includegraphics[scale=0.39]{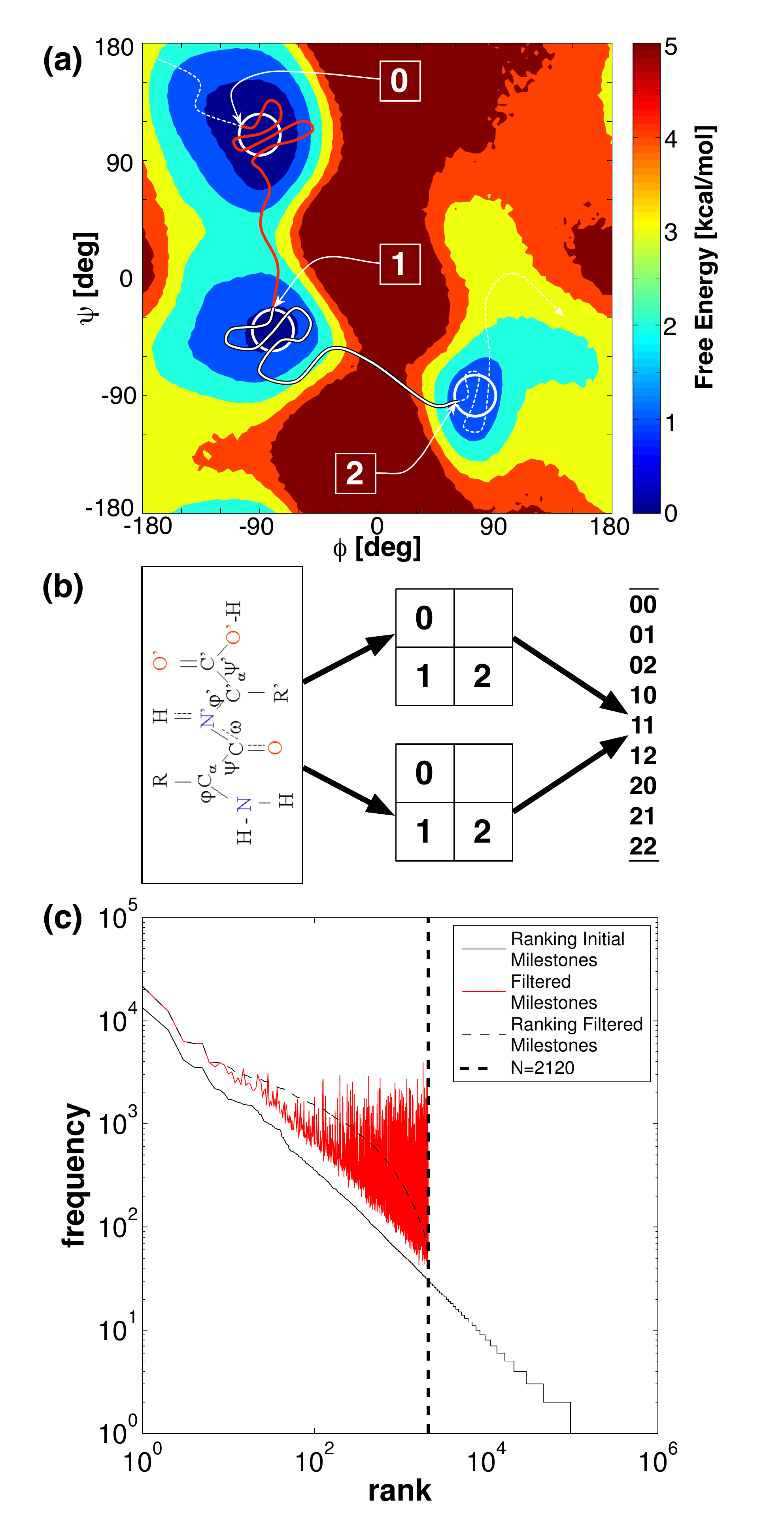}
\caption{\label{fig:diheMILE} Beta3s mini-protein: (a) Trajectory of the
  dihedral angles $(\phi,\psi)$ of a single amino acid crossing the
  three circular milestones $0$, $1$, and $2$. The part of the
  trajectory in red contributes to the statistics of the events
  $0\to1$ while the white double stroked portion contributes to the
  event $1\to2$. (b) Combinations of the dihedral
  milestones of two residues into chain milestones (strings). (c)
  Ranked frequencies of the dihedral milestones prior (black) and
  after (red) the filtering that yielded a trial set of $N=2120$
  milestones with a filter cutoff of 30. The unfiltered ranked
  frequencies follow the Zipf's law $1/r^{a}$ with exponent
  $a\approx0.76$.}
\end{figure}

To confirm that the target milestones in $\mathcal{M}^{(2)}$,
$\mathcal{M}^{(3)}$ and $\mathcal{M}^{(4)}$ were good core sets to
build an MSM (whereas others, including $\mathcal{M}^{(5)}$, were
not), we estimated the transition rate matrix entries
$k_{\alpha,\beta}$ of the corresponding MSMs using the procedure
described in Sec.~\ref{sec:MSM}. We then estimated the empirical first
passage time (FPT) distributions from the MD trajectory for the
transitions between the states in these MSMs. These were found to be
approximately exponential, with decay rates consistent with those
deduced from the matrix with entries $k_{\alpha,\beta}$ (data shown in
Fig. S1 of the supplementary materials \cite{SuppMatRef}). This indicates that the dynamics projected on
these states is approximately Markovian, and confirms that a low
metastability index is indeed a sign of Markovianity. Conversely, the
FPT distributions between milestones $C5$ and $P_{\text{II}}$ were
non-exponential, which is indicative of the non-Markovian character of
these transitions (see Fig. S1(a) in supplementary materials \cite{SuppMatRef}).
Interestingly, this result is consistent with the experimental
findings about the non-cooperativity of the alanine $P_{\text{II}}$
conformation in GGAGG peptides due to highly local hydration effects
\cite{Chen:2004aa}. It is also noteworthy that the transitions between
the target milestones occur on different time scales:
$\tau(P_{\text{II}}\to\alpha_{R})=0.48$~ns,
$\tau(\alpha_{R}\to P_{\text{II}})=0.73$~ns,
$\tau(P_{\text{II}}\to\alpha_{L})=9.5$~ns,
$\tau(\alpha_{R}\to\alpha_{L})=9.1$~ns,
$\tau(\alpha_{L}\to\alpha_{D})=8.5$~ns,
$\tau(\alpha_{D}\to\alpha_{L})=2.7$~ns.  This explains why different
sets of target milestones can be identified, similar to what we
observed in the illustrative example of
Sec.~\ref{sec:illustrativeex}. Note that the transition time
  scales between $P_{\text{II}}$ and $\alpha_{R}$ are consistent with
  the $\approx 1$ ns$^{-1}$ interconversion rates reported in earlier
  studies on a solvated alanine dipeptide
  \cite{Smith:1999aa,Bolhuis:2000aa,Oliveira:2007aa}. Note also that
the target milestone $\alpha_{L}$ plays the role of a hub for the
transitions involving the left quadrant of the Ramachandran plot with
the basin centered in $\alpha_{D}$. That is due to the fact that
direct transitions from the helical region $\alpha_{R}$ to the bottom
right quadrant are very rare events in a solvated alanine.

Finally, the MSM with the target milestones in $\mathcal{M}^{(4)}$ was
used to cluster the state space of the system. These results are shown
in Fig.~\ref{fig:GAG-commTScluster}. The committor probabilities
$q_{\alpha_R}(i)$ shown in Fig.~\ref{fig:GAG-commTScluster}(a) permit one
to assign each of the trial milestones to the target milestones it is
most likely to reach next, thereby partitioning the dihedral space into
basins (see Fig.~\ref{fig:GAG-commTScluster}(c)).  In
Fig.~\ref{fig:GAG-commTScluster}(b) the trial milestones are colored
according to the TSE index $\sigma(i)$ defined in~\eqref{eq:TSE}. The
red regions represent the trial milestones $S_{i}$ which are likely to
part of a transition state ensemble. Also shown in
Fig.~\ref{fig:GAG-commTScluster}(d) is the network of the MSM.

\section{Beta3s mini protein}
\label{sec:Beta3s}

As a last example we applied our method to analyze a 20 $\mu$s long MD
trajectory of the Beta3s peptide in implicit solvent at 330K, where
multiple folding/unfolding events were observed. The Beta3s is a 20
residue peptide which is known to assume a triple stranded
$\beta$-sheet fold
\cite{De-Alba:1999aa,Ferrara:2000aa,Rao:2004aa,Krivov:2008ac}.  The
simulation details of the MD data we analyzed here can be found in
Ref.~\cite{Muff:2008aa}. The 20 $\mathrm{\mu s}$ long trajectory is
composed of $10^{6}$ microstates saved at a lag-time $\tau=20$ ps.  In
the context of the GAG peptide (Sec.~\ref{sec:glyalagly}), the
torsional angles of the central alanine residue were shown to be
sufficiently good descriptors of the conformational space and were
used to define the trial milestones. This procedure, however, is not
applicable to a 20 residues mini protein such as Beta3s, whose
configurational space is much more complex than that of a
tri-peptide. For the initial definition of the trial set of milestones
several alternatives are in principle viable. Here we used the main
dihedral angles along the chain as starting point.

Specifically, the trial milestones were defined at the level of the
individual amino acids. For a protein of $R$ residues there are $R-2$
pairs of main dihedral angles $(\phi_{r},\psi_{r})$ along the chain,
where $r=2,...,(R-2)$ denotes the residue index. For each of the $R-2$
Ramachandran plots that are associated to a residue, three trial
milestones were defined as circles of radius 30 deg, and denoted as
milestone $0$, $1$, and $2$. The centers of these circles are located
at the minima of the free energy landscape obtained from the empirical
$(\phi_{r},\psi_{r})$ probability density of all the residues combined
(see the free energy contour plot in Fig.~\ref{fig:diheMILE}(a)) that
is estimated from the MD trajectory. The locations of the milestone
centers are $(-90,110)$ for state $0$ (beta-sheet region),
$(-80,-40)$ for the state $1$ (helix region), and $(75,-90)$ for state
$2$ (turn/loop region). In this way, the value of each pair of residue
$(\phi_{r},\psi_{r})$ along the MD trajectory is first mapped on a three-letter alphabet, and these letters are then combined
into a word, resulting in the representation
\begin{equation}
  S_{i}=s^{(2)}s^{(3)}\cdots s^{(R-2)}\label{eq:symbolMile}
\end{equation}
where $s^{(r)}\in(0,1,2)$ and $r$ the residue index. For example, the
folded state is represented as
``$000021000000210000$''. Fig.~\ref{fig:diheMILE}(b) gives a pictorial
representation of the construction of symbolic milestones. Since
Beta3s is a 20 residues mini-protein, the upper limit of accessible
trial milestones constructed this way is therefore
$3^{18}\approx 4\cdot10^{8}$.

\begin{figure}[t]
\includegraphics[scale=0.4]{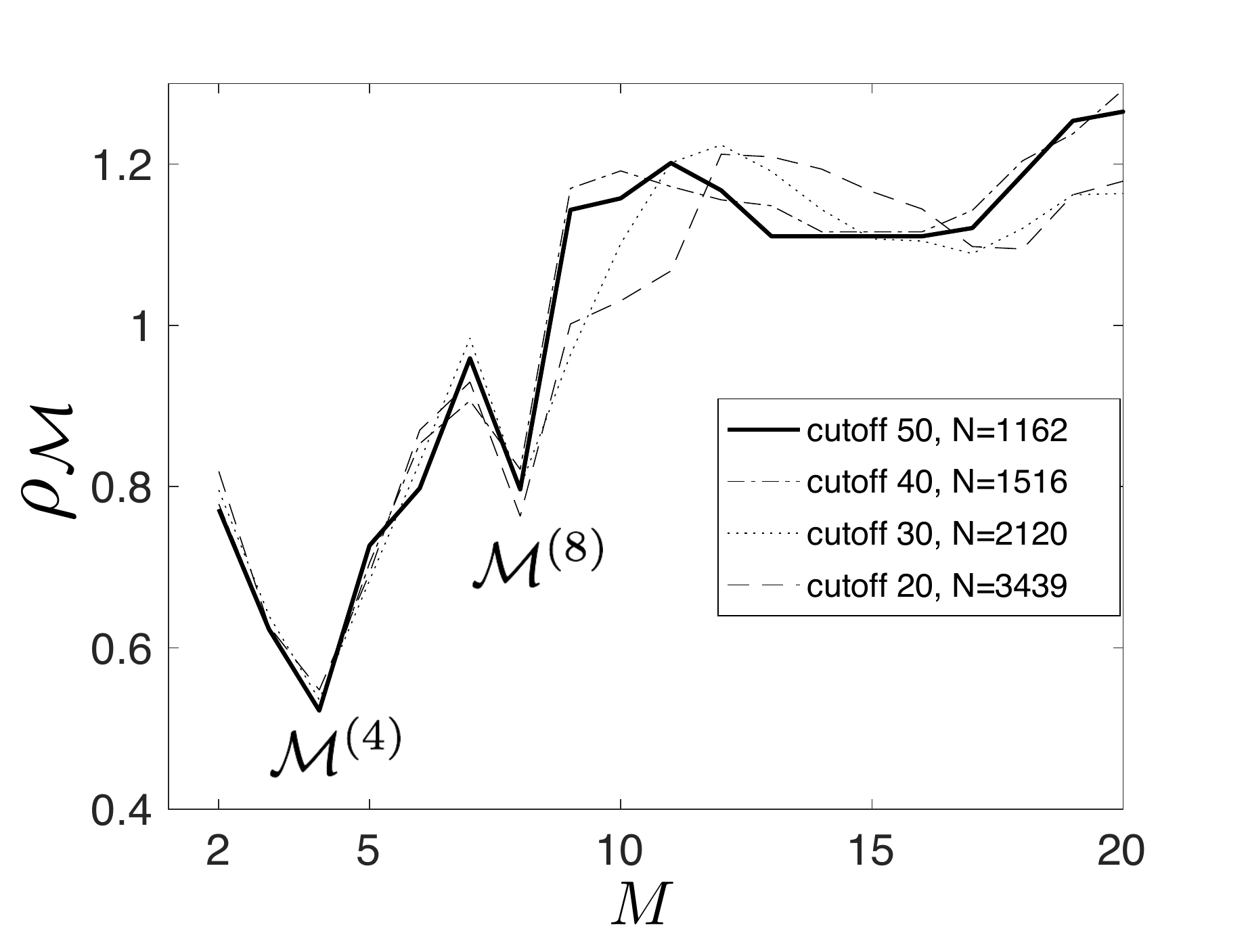}
\caption{\label{fig:metaIndex_Beta3s} Metastability index for the
  Beta3s peptide at 330K calculated using four different sets of trial
  milestones obtained with different filtering cutoff. In each case,
  the procedure identify the same target milestones with metastability
  index lower than 1..}
\end{figure}

\begin{figure*}
\includegraphics[scale=0.3]{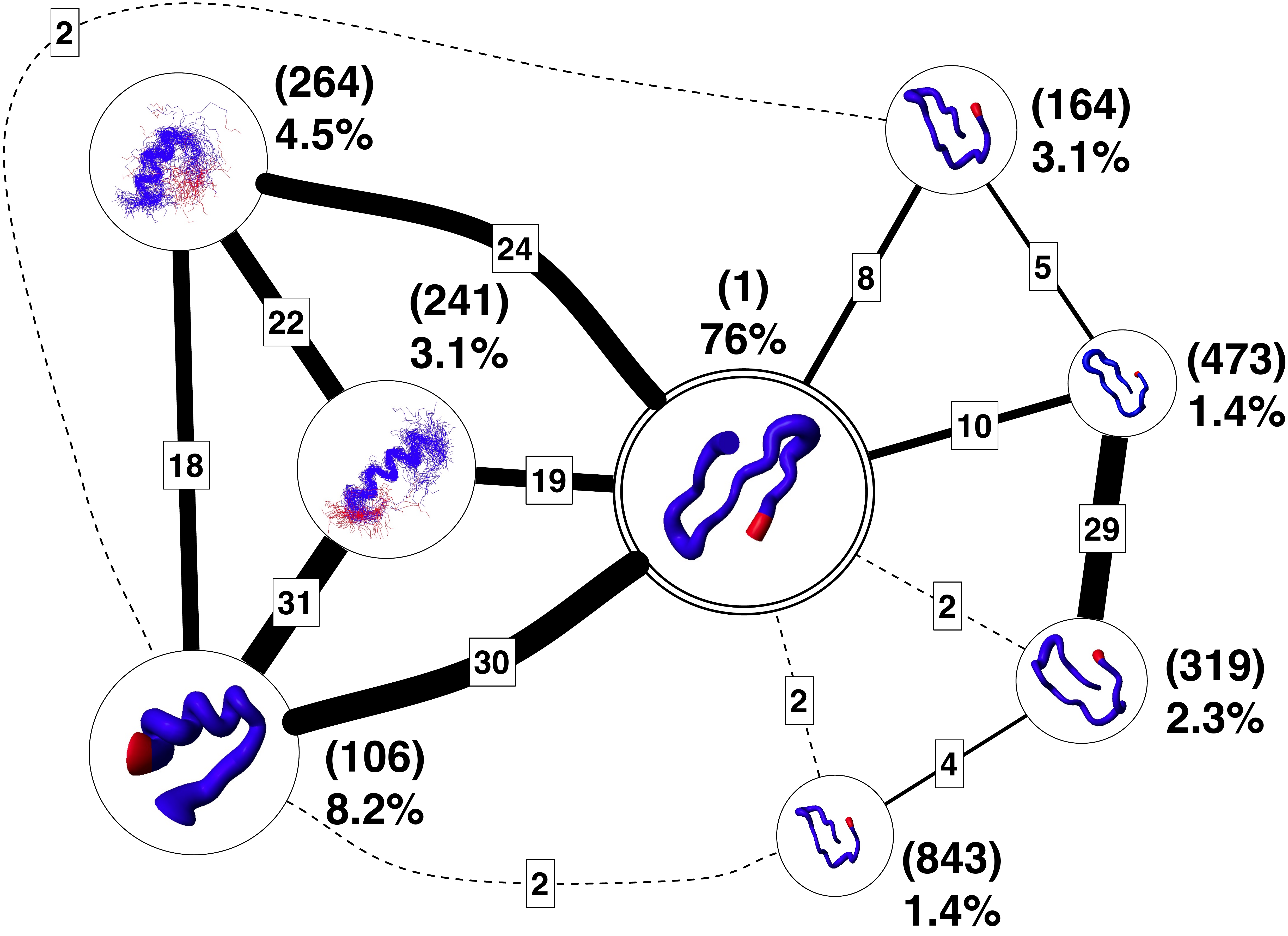}
\caption{\label{fig:reduNetwork}Network of the MSM for Beta3s based on
  $\mathcal{M}^{(8)}$. The nodes on the left side of the network
    have a high entropy while those on the right side have low energy.
    The node involving the triple stranded folded state lies at the
    center of a large basin in which energy and entropy compensate
    each other. The thickness of the edges between pairs of target
    milestones on the network is proportional to the total number of
    transitions observed between these milestones: these numbers are
    also reported in figure.  Each node of the network is
    represented along with the ensemble of structures associated to the most populated
    trial milestone it contains.}
\end{figure*}

In practice, not all these trial milestones were visited even once
along the trajectory, and some were visited much more often than
others. To avoid using trial milestones visited too few times, we
filtered out the milestones that were visited less than a threshold
value of times. The first step of dihedral milestoning gave a total
number of $N_{0}=469677$ dihedral strings, 373456 of which were
visited only once, which is about $\sim$37\% of the whole
set. Interestingly, the ranked distribution of the visting frequencies
follows a Zipf's law decay, namely a power law $\sim1/r^{a}$ with $r$
the rank and $a=0.76$ the exponent (see Figure \ref{fig:diheMILE}(c)).
Different frequency thresholds were used to filter out trial
milestones. Cutoffs corresponding to 20, 30, 40, and 50 reduced the
number of trial milestones to 3439, 2120, 1516, and 1162,
respectively. All these four sets of trial milestones were used to
process the MD trajectory and estimate the probabilities
$\Gamma_{\!i,j}$.  The metastability index $\rho_{\mathcal{M}}$ was
calculated for target sets $\mathcal{M}$ with cardinality in the range
$M=2,...,20$.  Fig.~\ref{fig:metaIndex_Beta3s} shows the result of
this calculation. Irrespective of the cutoffs used to define the set
of trial milestones, the resulting target sets were robust in the
range $M=2,...,8$ where at the low values of
$\rho_{\mathcal{M}}\lesssim1$.

To better understand the conformational dynamics of the Beta3s
  peptide, we used the tools introduced in
  Sec.~\ref{sec:diagno}. The committor probability $q_{\alpha}(i)$
  (obtained from \eqref{eq:12}) permits one to perform a soft partitioning
  of the trial milestones. Each target set milestone $B_{\alpha}$ is
  the center of a partition with statistical weight $\pi_{\alpha}$
  that is obtained from~\eqref{eq:13}, and from which the free energy
  $\Deltai G_\alpha$ is determined from~\eqref{eq:14}. This free
  energy can then be decomposed into energetic and entropic
  contributions according to~\eqref{eq:mileEi}
  and~\eqref{eq:TdS}. Here the effective energy of the system is
  identified as the sum of two contributions:
  $\bar E =E_{\rm charmm}+E_{\rm sasa}$ where $E_{\rm charmm}$ is the
  total potential energy in the CHARMM force field and $E_{\rm sasa}$
  is the solvent accessible term due to the implicit solvation
  model. Using~\eqref{eq:barE} we obtained $\bar{E}=-37.9\pm 11.1$
  kcal/mol as mean effective energy of the entire set of conformation
  sampled.

\begin{table}
\begin{tabular}{llccccc}
\hline 

\multicolumn{6}{c}{Markov State Model}\tabularnewline
id & Symbolic Milestone & $\pi_{\alpha}$ & $N_{\alpha}$ & $\Deltai
                                                          G_{\alpha}$
  & $\Deltai E_{\alpha}$ & $-T\Deltai S_{\alpha}$ \tabularnewline
& & {\footnotesize {[}\%{]}}& & {\tiny [kcal/mol]} & {\tiny [kcal/mol]} & {\tiny [kcal/mol]} \tabularnewline
\hline 
\hline 
\textbf{1} & \texttt{\footnotesize 000021000000210000}  & \textbf{76} & 48 & \textbf{0.2} & \textbf{-0.6} & \textbf{0.8} \tabularnewline
106 	   & \texttt{\footnotesize 111111111111100000 } & 8.2 & 42& 1.6 & 4.9 &   -3.3  \tabularnewline
164        & \texttt{\footnotesize 001211001000210010}  & 3.1 & 9 & 2.3 & -5.5 &  7.8  \tabularnewline
241        & \texttt{\footnotesize 111111111111111111 } & 3.1 & 37& 2.3 & 4.7 &   -2.5  \tabularnewline
264        & \texttt{\footnotesize 010001111111111111 } & 4.5 & 33& 2.0 & 5.1 &   -3.0  \tabularnewline
319 	   & \texttt{\footnotesize 001111000011111000 } & 2.3 & 19& 2.5 & -3.6 &  6.0  \tabularnewline
473 	   & \texttt{\footnotesize 001111000000200000 } & 1.4 & 23& 2.8 & -2.2 &  4.9  \tabularnewline
843        & \texttt{\footnotesize 001111000111210010}  & 1.4 & 5 & 2.8 & -2.8 &  5.7  \tabularnewline
\hline 

\end{tabular}

\caption{\label{tab:thermo-Mile} Beta3s mini-protein: Target milestones
  members of the set $\mathcal{M}^{(8)}$, along  with their thermodynamical parameters. }
\end{table}

We focused our analysis on the target set $\mathcal{M}^{(8)}$ that
resulted the largest metastable target set with
$\rho_{\mathcal{M}}<1$. The committor probability $q_{\alpha}(i)$ is
calculated from the time series of trial milestones, with
$\alpha\in \mathcal{M}^{(8)}$. Table \ref{tab:thermo-Mile} lists
  the statistical weights of the most populated states in the MSM
  build on $\mathcal{M}^{(8)}$. The mean effective energies
  $\Deltai E_{\alpha}$ are also reported in Table
  \ref{tab:thermo-Mile}, along with the entropy contributions
  $-T\Deltai S_{\alpha}$ to the free energy
  $\Deltai G_{\alpha}$. The basin centered around the target milestone
  associated with the folded state (id 1 in table
  \ref{tab:thermo-Mile}) has a statistical weight $\pi_1$ of about
  76\%, meaning that Beta3s spends about the 3/4 of the simulation
  time in this basin. As expected for a folded state, it has an
  energetic advantage ($\Deltai E_{1}=-0.6$ kcal/mol) which
  compensates its entropic disadvantage
  ($-T\Deltai S_{1}$=0.8 kcal/mol). The other basins
  identified in the target set $\mathcal{M}^{(8)}$ include helical and
  beta-curl type configurations. Interestingly, the helical basins (id
  106, 241, and 264 in Table~\ref{tab:thermo-Mile}) are entropically
  favored ($\Delta S_{\alpha}<0$ and $\Delta E_{\alpha}>0$) while
  the beta-curl states are energetically favored basins (id 164, 319,
  473, and 843 in Table~\ref{tab:thermo-Mile}) with
  $\Delta S_{\alpha}<0$ and $\Delta E_{\alpha}<0$. The statistical error on the mean effective energies $\Delta E_{\alpha}$ is estimated as $std(\Delta E_\alpha)/\sqrt{N_{\alpha}}$ where  $std(\Delta E_\alpha)$ is a standard deviation and $N_{\alpha}$ is the number of times the system enters a target milestone $B_{\alpha}$. The values of $N_{\alpha}$ are reported in Table~\ref{tab:thermo-Mile}. The statistical errors on $\Delta E_{\alpha}$ are reported in Fig. S4 of the supplementary materials \cite{SuppMatRef} along with the empirical probability density functions of the effective energy for each of the target milestones $B_{\alpha}$. While these errors are large, the information contained  in $\Deltai E_\alpha$ remains statistically significant.

Fig.~\ref{fig:reduNetwork} shows the network of the MSM built on the
target set $\mathcal{M}^{(8)}$. Notably, the folded state lies in
between the helical region, which is entropically favored, and the
beta-curl region, which is energetically favored. Except very rare
direct connections between the helical and beta-curl regions, most of
the transitions between the helix and beta-curl basins occur via the
folded state. Thus, as was remarked elsewhere \cite{Bowman:2010aa},
the folded state is not only the region of the configurational space
where energy and entropy compensate each other, but it also plays the
role of a dynamical hub in the network. This observation is
  confirmed by the high number of times the target milestone
  associated to the folded state is revisited ($N_{1}=48$ is the
  highest values observed, see $N_{\alpha}$ column in Table~\ref{tab:thermo-Mile}). 

  The high entropy character of the helical states is also reflected
  by the high connectivity within the helical region (left side of the
  network shown in Fig.~\ref{fig:reduNetwork}) and the comparable
    values of $N_{\alpha}$ with that of the folded state
    ($N_{106}=42$, $N_{241}=37$, and $N_{264} = 33$ in
    Table~\ref{tab:thermo-Mile}). On the contrary, the beta-curl
  states (right side of the network shown in
  Fig.~\ref{fig:reduNetwork}) are less connected ($N_{164}=9$,
  $N_{319}=19$, $N_{473}=23$, and $N_{843}=5$). This is consistent with
  the observation reported in \cite{Muff:2008aa} that these beta-curl
  states act as kinetic traps. 

  The probability distributions of FPT from the folded state to the
  other member of the target set and back are fairly well described by
  an exponential decay (see Fig. S2 and S3 in the supplementary materials \cite{SuppMatRef}). This indicates that
  the reduced dynamics of the Beta3s to a MSM built on the target set
  $\mathcal{M}^{(8)}$ is approximatively Markovian. Interestingly, the
  emerging picture from the network representation in
  Fig.~\ref{fig:reduNetwork} is qualitatively consistent with the
  simplified kinetic network representation proposed in Ref.~\cite{Krivov:2008ac}.

\begin{figure}
\label{fig:simplex}
\includegraphics[scale=0.4]{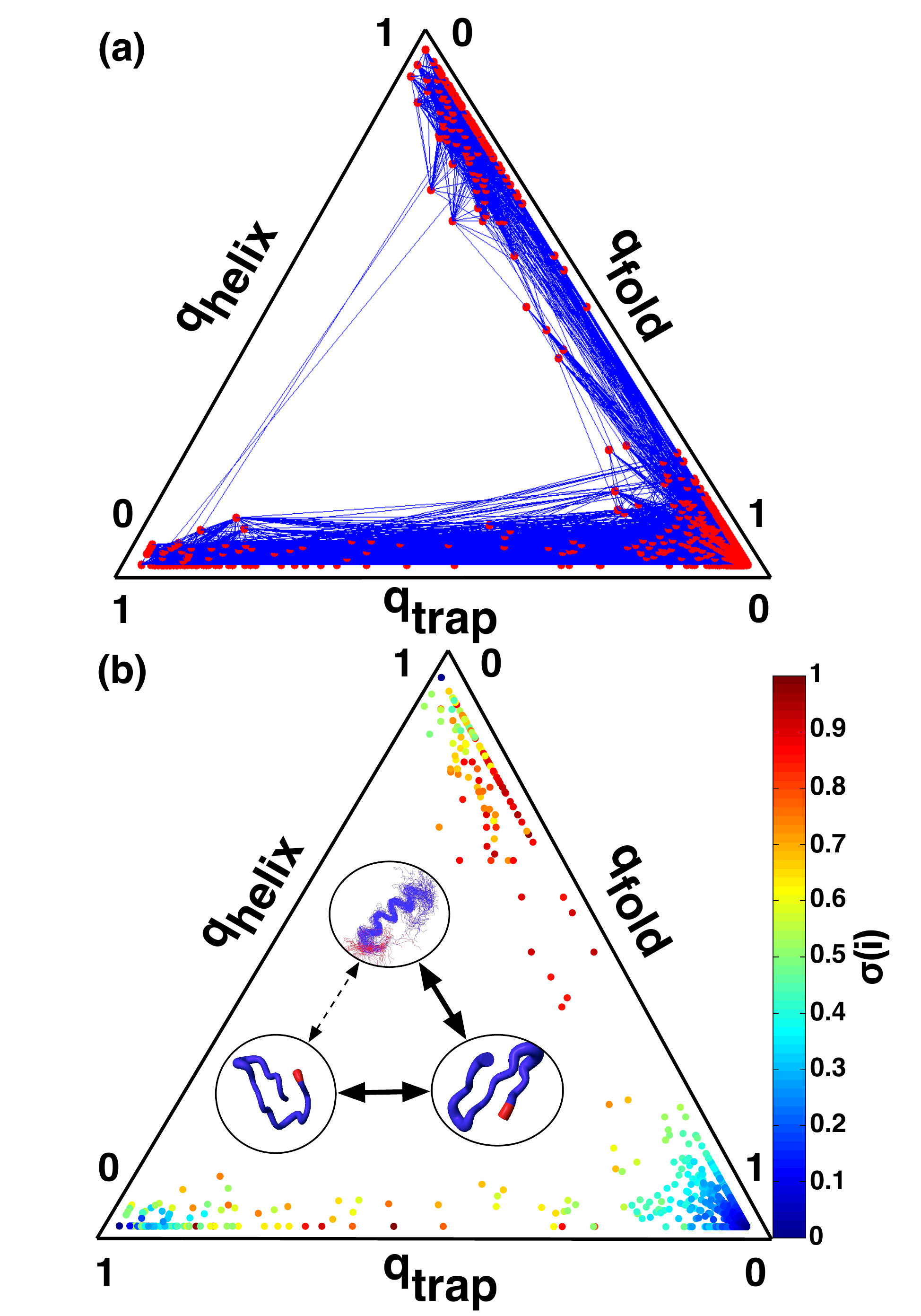}
\caption{Beta3s peptide: (a) Simplex
    representation of the aggregated committors such that
    $q_{{\rm fold}}=q_{1}$, $q_{{\rm helix}}=q_{106}+q_{241}+q_{264}$
    and $q_{{\rm trap}}=q_{164}+q_{319}+q_{473}+q_{843}$ (these
    indices are the same as those used in
    Fig.~\ref{fig:reduNetwork}). The continuous blue line represent
    the time series $q_ {{\rm fold}}(i_*(t))$,
    $q_{{\rm helix}}(i_*(t))$ and $q_{{\rm trap}}(i_*(t))$ Only 4
    transition events connect directly the helix and beta regions. 
  (b) Same representation with nodes colored according 1 to the TS
  index $\sigma(i)$ defined in~\eqref{eq:TSE}.}
\end{figure}

The committor probability $q_{\alpha}(i)$ calculated for the target
set $\mathcal{M}^{(8)}$ allows one to project the configurational space
onto a simplex. Three aggregated committor probabilities are defined
as $q_{{\rm fold}}=q_{1}$ for the folded state,
$q_{{\rm helix}}=q_{106}+q_{241}+q_{264}$ for the helix state, and
$q_{{\rm trap}}=q_{164}+q_{319}+q_{473}+q_{843}$ for the beta-curl
state (see Table \ref{tab:thermo-Mile} for the state indexes). By
definition of committor probability one has
$q_{\rm fold}+q_{\rm helix}+q_{\rm trap}=1$. Fig.~\ref{fig:simplex}
shows the simplex representation of the aggregated committor
probabilities in a triangular plot. Fig.~\ref{fig:simplex}(a) shows
the time series of the committor probabilities
$q_{\rm fold}(i_*(t)), q_{\rm helix}(i_*(t)),q_{\rm trap}(i_*(t))$ projected onto a
triangular plot. Interestingly, the conformers are clustered around
the three vertexes of the triangular plot. The blue lines represent
direct transitions between conformer and only 4 direct transitions
link the helix channel to the trap channel. Fig.~\ref{fig:simplex}(b)
shows a triangular scatter plot of the committor probabilities with
the points colored according to their TSE index $\sigma(i)$  defined
in \eqref{eq:TSE}. Red points correspond to milestones belonging to
transition state ensembles. For instance, the channel connecting the
helix to the fold basin is composed by milestones $S_{i}$ such that
$q_{\rm fold}\approx q_{\rm helix}\approx 0.5$. Similarly, along the
channel Trap$\leftrightarrow$Fold one has milestones $S_{i}$ with
$q_{\rm fold}\approx q_{\rm trap}\approx 0.5$. The simplex
  representation in Fig.~\ref{fig:simplex} corroborates the
  observation that the folded state acts as a kinetic hub
  \cite{Muff:2008aa}, and it also suggests that Beta3s oscillates
  between two extreme conformational regions: a low energy one versus
  a high entropy one.  

\section{Conclusions}
\label{sec:conclu}

Despite their growing popularity to analyze and interpret molecular
dynamics (MD) simulations, Markov State Models (MSMs) remain somewhat
tricky to use and justify. How to pick the states over which to build
the MSM and how to assess its accuracy are nontrivial issues that are
related to the degree to which the dynamical coarse-graining over these
states preserves the Markovianity of the original system. Typically,
states in MSMs are picked in a somewhat \textit{ad~hoc} fashion, and
their quality is assessed \textit{a~posteriori} via Markovianity
tests. In this paper, we propose a strategy that may help systematize
the operation of MSM building. It is based on the observation that, in
metastable systems, the states in an MSM should be hubs to which the
trajectory has a high probability to return often, and between which
it seldom transitions. These intuitive properties can be captured via
a metastability index that measures how good a set of target
milestones chosen among trial ones is: the smaller its metastability
index, the better the target milestones it contains are as hubs. This
metastability index is not a new concept: it was introduced in a
closely related form in potential theoretic approaches to
metastability. Our main goal in this paper was to show that
this index can also be turned into a practical computational tool for
the identification of good target milestones upon which to build
accurate MSMs. As we shown here the procedure is simple to use, it
does not require to make Markovian assumptions at the intermediate
stages of the construction (in particular, the jumps between the trial
milestones could be correlated and non-Markovian), and it has the
added advantage that it allows one to cluster the system's state space
\textit{a~posteriori} and identify the transition state ensembles
(TSEs) between the target milestones. Here, these features were
illustrated on a collections of examples of increasing complexity. In
particular, our last test case involving the Beta3s mini-protein,
showed that the method can reveal interesting properties of the
folding landscape of proteins, including the presence of kinetic traps
and misfolded states on the way to the native state, and how these
features affects the folding and unfolding pathways of the protein. We
certainly hope that the technique will be similarly useful to analyze
the dynamics of other proteins and macromolecules.

\section*{Acknowledgments}
We thank Amedeo Caflisch for sharing the MD trajectories of the Beta3s
peptide. We also thank Jianfeng Lu for helpful discussions.

\appendix

\section{Derivation of \eqref{eq:GammaMFPT}}

In this section a proof of~\eqref{eq:GammaMFPT} is given in case of a
Markov process with generator $L_{i, j}$ and transition probability
$P_{i, j} = L_{i,j}/\sum_{k\not=i} L_{i,k}$, $i\not=j$,
$P_{i,i}=0$. To begin, recall that the MFPT $\tau_{i,j}$ from state
$i$ to state $j$ is the solution to
\begin{equation}
  \sum_{k =1}^N L_{i, k} \tau_{k,j} = - 1, \quad \text{for} \ \
  i\not=j, \quad 
  \text{and}\quad
  \tau_{j,j} = 0
  \label{eq:MFPTmat}
\end{equation}
while  the MRT is given by
\begin{equation}
  \tau_{i} = (\sum_{j \neq i} L_{i, j})^{- 1} + \sum_{k=1}^N P_{i, k} \tau_{k,i}
  \label{eq:MRT}
\end{equation}
where the first term at the right hand-side accounts for the mean
time to exit $i$ to some state $j\not= i$, and the second for the mean
time to come back from any such state $j\not=i$ to $i$.
We are interested to find an expression for the probabilities
$\Gamma_{\!i,j}$ as a function of the MFPT and MRT. Since
$\Gamma_{\!i,j} = \sum_k P_{i,k} q_{i, j} (k)$, we can equivalently
express the committor function $q_{i, j} (k)$ in terms of MFPT and the
MRT. We claim that
\begin{equation}
  q_{i, j} (k) = \frac{\tau_{k,i} - \tau_{k,j}+ 
    \tau_{i,j}}{\tau_{i,j} + \tau_{j,i}} \label{eq:CommMFPT}
\end{equation}
Let us check that this relation holds by verifying that the function
$q_{i,j}(k)$ defined this way satisfies $q_{i, j} (i) = 0$,
$q_{i, j} (j) = 1$ and $\sum_{l \in S} L_{k,l} q_{i, j} (l) = 0$ for
$k \notin \{i, j\}$. The two boundary conditions are trivially
verified and the third condition is
\begin{equation}
  \begin{aligned}
    & \sum_{l =1}^N L_{k,l} q_{i, j} (l) \\
    & \quad = \frac{1}{\tau_{i,j} + \tau_{j,i}} \sum_{k,l=1}^N L_{k,l}
    (\tau_{l,i} - \tau_{l,j} + \tau_{i,j})\\
    &\quad =0
  \end{aligned}
\label{eq:CondL}
\end{equation}
where we have used~\eqref{eq:MFPTmat}. We can now calculate
$\Gamma_{\!i,j}$ using \eqref{eq:CommMFPT} and \eqref{eq:TransMat}
\begin{equation}
  \begin{aligned}
    \Gamma_{\!i,j} & =  \sum_{k =1}^N
    P_{i, k} q_{i, j} (k)\\
    & =  \frac{(\sum_{k \neq i} L_{i, k})^{- 1}}{\tau_{i,j}
    + \tau_{j,i}}  \sum_{k =1}^N L_{i, k} 
    (\tau_{k,i} - \tau_{k,j} + \tau_{i,j})\\
    & = \frac{(\sum_{k \neq i} L_{i, k})^{- 1}}{\tau_{i,j}
    + \tau_{j,i}}  \left( \sum_{k=1}^N L_{i, k}
    \tau_{k,i} + 1 \right)\\
    & =  \frac{\tau_{i}}{\tau_{i,j} + \tau_{j,i}}
  \end{aligned} \label{eq:PcommAB}
\end{equation}
where we used~\eqref{eq:MFPTmat} and \eqref{eq:MRT}. Thus, for any
pair of states $(i,j)$, the probability $\Gamma_{\! i,j}$ can be computed
from the MFPTs and MRTs only. 



%

\end{document}